%% file: main.tex
\documentclass[fleqn,usenatbib]{mnras}
\usepackage{newtxtext,newtxmath}
\usepackage{lineno}
\usepackage[T1]{fontenc}
\usepackage{ae,aecompl}
\usepackage{multirow}

\usepackage{graphicx}	% Including figure files
\usepackage{amsmath}	% Advanced maths commands
\usepackage{upgreek}
\newcommand{\tess}{\emph{TESS}\xspace}
\newcommand{\gaia}{\emph{Gaia}\xspace}

\providecommand{\bjdtdb}{\ensuremath{\rm {BJD_{TDB}}}}

\providecommand{\meh}{\ensuremath{\left[{\rm m}/{\rm H}\right]}}
\providecommand{\feh}{\ensuremath{\left[{\rm Fe}/{\rm H}\right]}}
\providecommand{\teff}{\ensuremath{T_{\rm eff}}}
\providecommand{\logg}{\ensuremath{\log g}}
\providecommand{\vsini}{\ensuremath{v\sin i}}

\providecommand{\msun}{\ensuremath{M_\odot}}
\providecommand{\rsun}{\ensuremath{R_\odot}}

\providecommand{\rj}{\ensuremath{R_{\rm Jup}}}
\providecommand{\mj}{\ensuremath{M_{\rm Jup}}}

\newcommand{\ms}{\,m\,s$^{-1}$}
\newcommand{\kms}{\,km\,s$^{-1}$}

\newcommand{\tic}{TIC 436873727\xspace}

\newcommand{\toi}{TOI-4641\xspace}
\newcommand{\toib}{TOI-4641b\xspace}

\newcommand{\pmRA}{$61.447\pm0.027$}
\newcommand{\pmDEC}{$-20.020\pm0.024$}
\newcommand{\parallax}{$11.409\pm0.026$}
\newcommand{\vsiniSPC}{$91.60\pm0.50$}
\newcommand{\vsiniLSD}{$86.3\pm1.0$}
\newcommand{\loggfit}{$4.11^{+0.69}_{-0.69}$}
\newcommand{\tefffit}{$6560^{+300}_{-340}$}
\newcommand{\metfit}{$-0.09^{+0.11}_{-0.18}$}
\newcommand{\vsinifit}{86.3$^{+1.00}_{-0.99}$}
\newcommand{\mstar}{$1.41^{+0.068}_{-0.059}$}
\newcommand{\rstar}{$1.72^{+0.041}_{-0.043}$}
\newcommand{\lstar}{$4.95^{+0.94}_{-1.21}$}
\newcommand{\age}{2.69$^{+0.81}_{-1.14}$}
\newcommand{\dist}{87.60$^{+0.20}_{-0.20}$}
\newcommand{\macroturb}{4.00$^{+0.62}_{-0.66}$}

\newcommand{\per}{22.093410$^{+0.000051}_{-0.000047}$}
\newcommand{\midt}{2459510.82759$^{+0.00091}_{-0.00100}$}

\newcommand{\plrad}{0.730$^{+0.026}_{-0.028}$}
\newcommand{\rprs}{0.042$^{+0.0009}_{-0.0009}$}
\newcommand{\incl}{87.90$^{+0.088}_{-0.084}$}

\newcommand{\bimp}{0.780$^{+0.014}_{-0.015}$}
\newcommand{\semimaj}{0.173$^{+0.015}_{-0.015}$}

\newcommand{\lam}{$1.41^{+0.76}_{-0.76}$}
\newcommand{\aor}{$21.53^{+0.51}_{-0.52}$}
\newcommand{\tdur}{$0.2266^{+0.0029}_{-0.0030}$}

\newcommand{\CfA}{Center for Astrophysics \textbar \ Harvard \& Smithsonian, 60 Garden Street, Cambridge, MA 02138, USA}
\newcommand{\USQ}{University of Southern Queensland, Centre for Astrophysics, West Street, Toowoomba, QLD 4350 Australia}
\newcommand{\FlatironCCA}{Center for Computational Astrophysics, Flatiron Institute, 162 Fifth Avenue, New York, NY 10010, USA}

\newcommand{\MITKavli}{Department of Physics and Kavli Institute for Astrophysics and Space Research, Massachusetts Institute of Technology, Cambridge, MA 02139, USA}
\newcommand{\DTU}{DTU Space,  Technical University of Denmark, Elektrovej 328, DK-2800 Kgs. Lyngby, Denmark}

\usepackage{xspace}
\usepackage{scalerel}
\usepackage{tikz}
\usetikzlibrary{svg.path}

\definecolor{orcidlogocol}{HTML}{A6CE39}
\tikzset{
  orcidlogo/.pic={
    \fill[orcidlogocol] svg{M256,128c0,70.7-57.3,128-128,128C57.3,256,0,198.7,0,128C0,57.3,57.3,0,128,0C198.7,0,256,57.3,256,128z};
    \fill[white] svg{M86.3,186.2H70.9V79.1h15.4v48.4V186.2z}
                 svg{M108.9,79.1h41.6c39.6,0,57,28.3,57,53.6c0,27.5-21.5,53.6-56.8,53.6h-41.8V79.1z M124.3,172.4h24.5c34.9,0,42.9-26.5,42.9-39.7c0-21.5-13.7-39.7-43.7-39.7h-23.7V172.4z}
                 svg{M88.7,56.8c0,5.5-4.5,10.1-10.1,10.1c-5.6,0-10.1-4.6-10.1-10.1c0-5.6,4.5-10.1,10.1-10.1C84.2,46.7,88.7,51.3,88.7,56.8z};
  }
}

\newcommand\orcid[1]{\href{https://orcid.org/#1}{\mbox{\scalerel*{
\begin{tikzpicture}[yscale=-1,transform shape]
\pic{orcidlogo};
\end{tikzpicture}
}{|}}}}
\title[TOI-4641b]{TOI-4641b: An Aligned Warm Jupiter Orbiting a Bright (V=7.5) Rapidly Rotating F-star}

\input{authors.tex}

\date{Accepted 2023 December 05. Received 2023 December 05; in original form 2023 September 17}

\pubyear{2023}

\begin{document}
\label{firstpage}
\pagerange{\pageref{firstpage}--\pageref{lastpage}}
\maketitle

\clearpage
\begin{abstract}
We report the discovery of \toib, a warm Jupiter transiting a rapidly rotating F-type star with a stellar effective temperature of 6560 K. The planet has a radius of 0.73~\rj, a mass smaller than 3.87~\mj~$(3\sigma)$,
and a period of 22.09 days. It is orbiting a bright star (V=7.5 mag) on a circular orbit with a radius and mass of 1.73 \rsun\, and 1.41 \msun. Follow-up ground-based photometry was obtained using the \textit{Tierras} Observatory. Two transits were also observed with the Tillinghast Reflector Echelle Spectrograph (TRES), revealing the star to have a low projected spin-orbit angle ($\lambda$=\lam$\degr$). Such obliquity measurements for stars with warm Jupiters are relatively few, and may shed light on the formation of warm Jupiters. Among the known planets orbiting hot and rapidly-rotating stars, TOI-4641b is one of the longest-period planets to be thoroughly characterized. Unlike hot Jupiters around hot stars which are more often misaligned, the warm Jupiter TOI-4641b is found in a well-aligned orbit. Future exploration of this parameter space can add one more dimension to the star-planet orbital obliquity distribution that has been well-sampled for hot Jupiters.
\end{abstract}

\begin{keywords}
%Extrasolar gaseous giant planets (509) --- Radial velocity (1332) --- Transit photometry (1709)
%https://academic.oup.com/DocumentLibrary/mnras/keywords.pdf
exoplanets - techniques: radial velocities - techniques: spectroscopic - techniques: photometric  - methods: observational
%https://static.primary.prod.gcms.the-infra.com/static/site/mras/document/Updated+keyword+list+(Aug%202022).pdf?node=2777349d087f40bd3e40
\end{keywords}

%%%%%%%%%%%%%%%%%%%%%%%%%%%%%%%%%%%%%%%%%%%%%%%%%%
% \tableofcontents

\section{Introduction} \label{sec:Intro}

Many hot Jupiters are thought to have formed at least a few astronomical units away from their stars and migrated inward via dynamical interactions. One possible migration mechanism is high-eccentricity tidal migration, whereby their orbits gradually circularize with each periastron passage due to dissipative tidal interactions with their host stars \citep[e.g.][]{1996Sci...274..954R}. They may instead have formed in-situ \citep[e.g.][]{batygin2012} or undergone disk-driven migration soon after their formation \citep[e.g.][]{1996Natur.380..606L}. Gas giants in somewhat longer period orbits, where tidal effects are negligible, provide useful laboratories for the study of the migration processes that produce hot Jupiters \citep{2018DawsonJohnson}. 

Spin-orbit angle, the angle between the orbital normal and the spin-axis of the host star, serves as a fossil record of past dynamical interactions experienced by the system. In particular, warm Jupiters, planets with $a/R_{\star} > 10$ and $R_{p} > 8\, R_{e}$, offer an opportunity to study the primordial obliquities of their host stars \citep{albrecht2022} without having to account for planet-star tidal interactions that may have modified the orbital architecture. 

There exists clear trends in the hot Jupiter spin-orbit obliquity distribution as a function of stellar temperature \citep[e.g.][]{albrecht2012b}. The dependence between the observed obliquity distribution and stellar temperature for longer period planets is less clear due to observational biases that make such observations more difficult. Close-in giant planets around cool stars (\teff\, $< 6250$ K) are generally observed to have well-aligned orbits \citep{albrecht2022}. In contrast, more distantly orbiting giant planets about cool stars with $a/R_{\star} > 10$ show a wider range of spin-orbit angles, as in systems like WASP-8 \citep{WASP8_2010Queloz,WASP8_2017}, Kepler-420 \citep{Kepler420_2014Santerne}, and HD 80606 \citep{HD806062009Winn, HD806062009Pont}. Such a dependence is expected if the close-in planet obliquity distribution is strongly shaped by planet-star tidal interactions. 

Short-period Jovian planets around early-type stars exhibit a wide distribution of orbital obliquities. Few long-period planets around hot stars have had their spin-orbit angles mapped. Kepler-448b is the only Jovian planet around an early-type star with an spin-orbit angle measured spectroscopically found to reside in a well-aligned orbit \citep{2015A&A...579A..55B,2017AJ....154..137J}. 

The lack of well characterized long period planets around early-type stars prohibits informative tests on the mechanisms thought to induce misalignments in planet orbits. Early-type stars with radiative envelopes experience weaker planet-star tidal interactions and offer the opportunity of exploring the primordial period obliquity relationship for giant planets. Each mechanism that induces planet-star misalignments have their own expected dependencies on orbital distance. Warps in the protoplanet disk lead to preferentially longer period planets being found in misaligned orbits \citep{1993ApJ...408..337H,2017A&A...604A..88W}. The spin-axis of stars hotter than the Kraft break \citep{kraftbreak1967} may also evolve over time by itself. Gravity-wave instabilities induced at the radiative-convective boundary of early-type stars may result in their outer envelopes changing in spin-axis over time \citep{2012ApJ...758L...6R}, leading to a wide range of spin-orbit obliquites for planets around such stars. Such instabilities are not dependent on the planetary system, and as such no period-spin-orbit obliquity dependencies should be expected. 

In this paper we report the planetary confirmation of \toib, a warm Jupiter on a 22-day orbit that is well aligned with the equatorial plane of a rapidly rotating F-star. In Section \ref{sec:Observations}, we describe the photometric data from 
\tess and the \textit{Tierras} Observatory, high-resolution speckle imaging, and TRES spectroscopic observations to measure the stellar obliquity. Section \ref{sec:global} describes the global modeling of the system and Section \ref{sec:variability} describes the stellar variability. We conclude with a discussion in Section \ref{sec:discussion}.

%%%%%%%%%%%%%%%%%%%%%%%%%%%%%%%%%%%%%%%%%%%%%%%%%%%%%%%%%%%%%%%%%%%%%%%%%
\section{Observations} \label{sec:Observations}

\subsection{Photometric Observations} \label{subsec:photometry}
\subsubsection{TESS Photometry} \label{subsec:TESS}
\begin{figure*}
\begin{center}
% \epsscale{1}
\includegraphics[width=\textwidth]{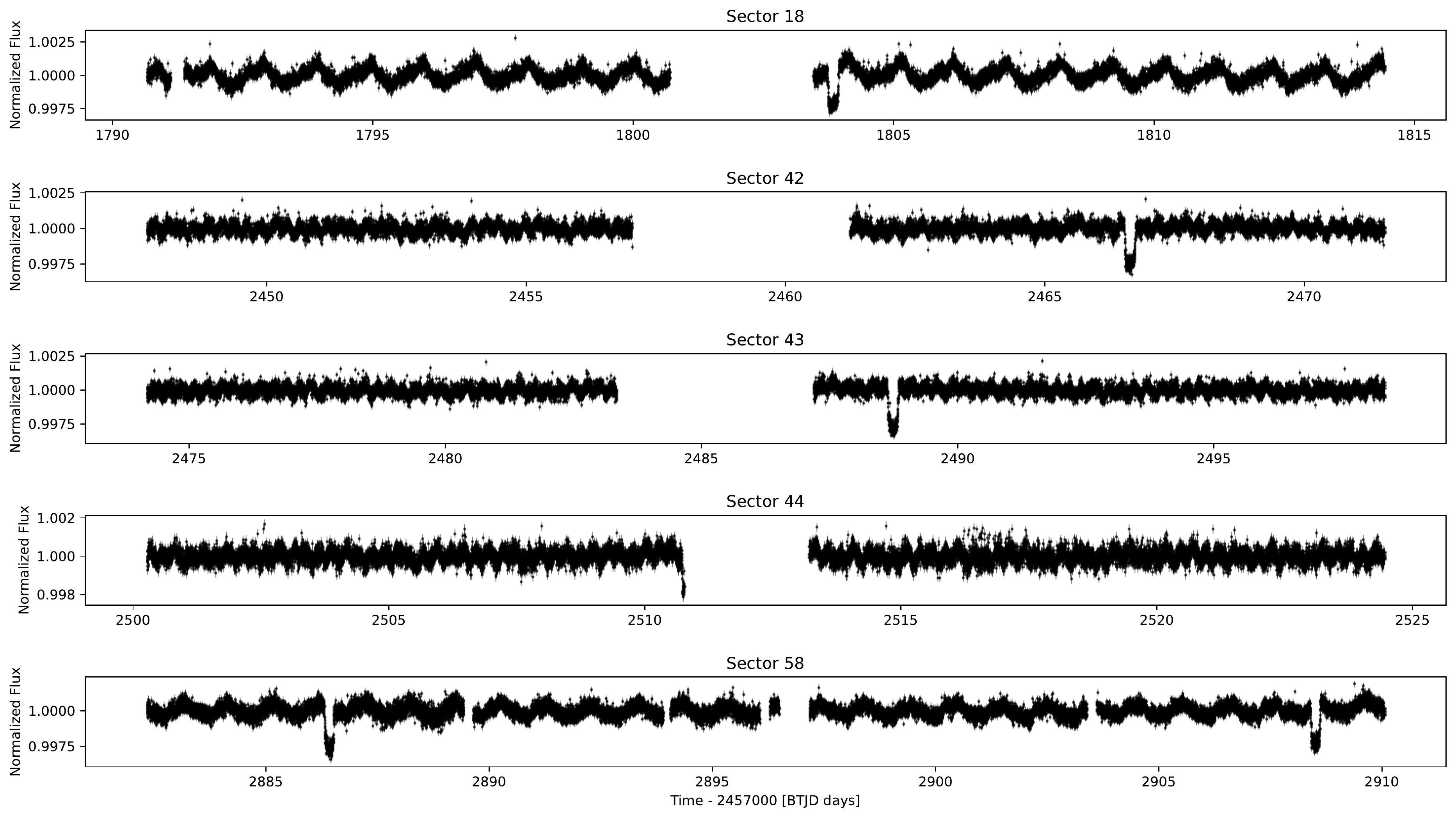}
\caption{Per-sector Normalized TESS PDCSAP light curves for TOI-4641. The target star was observed over five TESS sectors. Because the Sector-44 dataset covers only one part of one transit, this dataset was not included in our global modeling of the planetary properties. Significant stellar variability is seen within each sector of the TESS observations, as discussed in Section~\ref{sec:variability}.} 
\label{fig:raw_tess}
% \end{minipage}
\end{center}
\end{figure*}

%SPOC sectors 18, 43, 44, 58 
The Transiting Exoplanet Survey Satellite (\textit{TESS}; \citealt{ricker2015}) is an all-sky survey searching for transiting exoplanets around nearby bright host stars. The satellite uses four cameras to stare at 24 $\times$ 96 degree sectors of the sky for approximately 27 days at a time. \toi (\tic) was observed by \textit{TESS} during Sector 18  of the primary mission and then again in Sectors 42, 43, 44, and 58 of the extended mission, in all cases with 2-minute cadence.  The data were processed by the NASA Science Processing Operations Center pipeline (SPOC; \citealt{jenkins2016}) and the light curves were downloaded from the Mikulski Archive for Space Telescopes (MAST)\footnote{https://mast.stsci.edu/portal/Mashup/Clients/Mast/Portal.html} using the  \textit{Lightkurve} package \citep{lightkurve}. The Presearch Data Conditioning Simple Aperture Photometry (PDCSAP; \citep{2012PASP..124..985S,2014PASP..126..100S,2012PASP..124.1000S}) light curves were employed in our analysis and are plotted in Figure \ref{fig:raw_tess}. Sector 44 was excluded from our planet analysis because the transit occurred at the edge of a gap in the lightcurve.

A candidate exoplanet orbiting TOI-4641 with a period of 22.1d was identified in light curves including data through Sector 43 in both SPOC \citep{jenkins2016} and QLP \citep{Huang2020,Kunimoto2022} pipelines. The SPOC performed a transit search with an adaptive, noise-compensating matched filter \citep{Jenkins2002,Jenkins2010,Jenkins2020}, producing a Threshold Crossing Event (TCE) for which an initial limb-darkened transit model was fitted \citep{Li2019} and a suite of diagnostic tests were conducted to help assess the planetary nature of the signal \citep{Twicken2018}. The QLP performed its transit search with the Box Least Squares Algorithm \citep{Kovacs2002}. The transit signature passed all SPOC data validation diagnostic tests, and the TESS Science Office issued an alert \citep{Guerrero2021} for TOI 4641.01 on 19 November 2021. The difference image centroid offsets localized the transit source for TOI 4641.01 within $2.4\pm2.5$ arcsec; all TIC v8 \citep{stassun2019} objects other than TOI-4641 were excluded as potential sources of the transit signature.

\subsubsection{Ground-based Photometry} \label{subsec:Tierras}

To check the field for nearby eclipsing binaries that could potentially be contaminating the \tess photometry and to confirm the transit was on the target star, we used the {\tt TESS Transit Finder}, a customized version of the {\tt Tapir} software package \citep{jensen2013}, to schedule ground-based transit observations. Follow-up observations were done using the \textit{Tierras} Observatory \citep{tierras} at the Fred Lawrence Whipple Observatory atop Mount Hopkins in Arizona, USA. \textit{Tierras} is a 1.3m telescope with a Teledyne e2v 4K $\times$ 4K NIR-optimized deep-depletion CCD. It has a 0.48 $\times$ 0.25 degree field-of-view and a $0.43$ arcsecond per pixel scale. The camera was designed to have a custom narrow (40 nm full width at half maximum) bandpass filter centered around 863.5 nm to minimize precipitable water vapor errors known to limit ground-based photometry of M dwarfs. 

\begin{figure}[]
    \centering
    \includegraphics[width=0.46\textwidth]{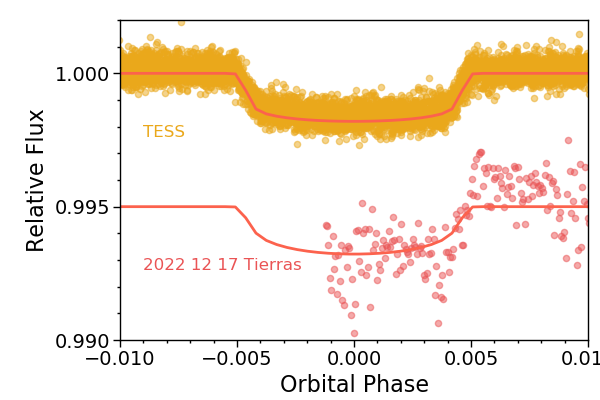}
    \caption{\textbf{Top:} Phase folded \textit{TESS} light curves of 5 transits from Sectors 18, 42, 43, and 58. The best fit model from the global modeling analysis is illustrated as a red line. \textbf{Bottom:} Follow-up observations from the \textit{Tierras} Observatory, capturing the egress event on UT December 17, 2022.}\label{fig:lc}
\end{figure}

The data were reduced using a custom pipeline based on similar procedures as outlined in \cite{irwin2015} and aperture photometry was performed using AstroImageJ \citep[AIJ,][]{aij}. Two partial transits were observed on UT December 17, 2022 (egress) and UT January 30, 2023 (ingress). A full transit was observed on UT January 8, 2023. The transits on the nights UT December 17, 2022 and UT January 8, 2023 coincided with the spectroscopic observations described in more detail in Section \ref{subsec:TRES}. We extracted the photometry using an aperture radius of 12 pixels  (5.16") from the data on UT December 17, 2022 to determine that the target star was the source of the transit events and we excluded any star more distant than 5 arseconds as the source of the dips observed by \tess. That light curve is shown in Figure \ref{fig:lc} along with the phase folded \tess\,light curve. We note $\sim 6$ hour variation in the ground-based light curve at the $\sim 1$ mmag level. The light curve variability is discussed in more detail in Section \ref{sec:variability}. Data from the nights of UT January 8 and January 30, 2023 were not of sufficient quality to add any value to the fit due to poor observing conditions and experimental exposure times leading to occasional saturation and were not included in the global analysis. The photometric analysis files from AIJ are all available publicly on the Exoplanet Follow-up Observing Program (ExoFOP)\footnote{https://exofop.ipac.caltech.edu/tess/} website.

%TESS Follow-up Observing Program\footnote{https://tess.mit.edu/followup} Sub Group 1 \citep[TFOP;][]{collins:2019}. 

\subsubsection{High-resolution Imaging} \label{subsec:imaging}

\begin{figure}
  \centering    \includegraphics[width=.9\linewidth,height=175pt]{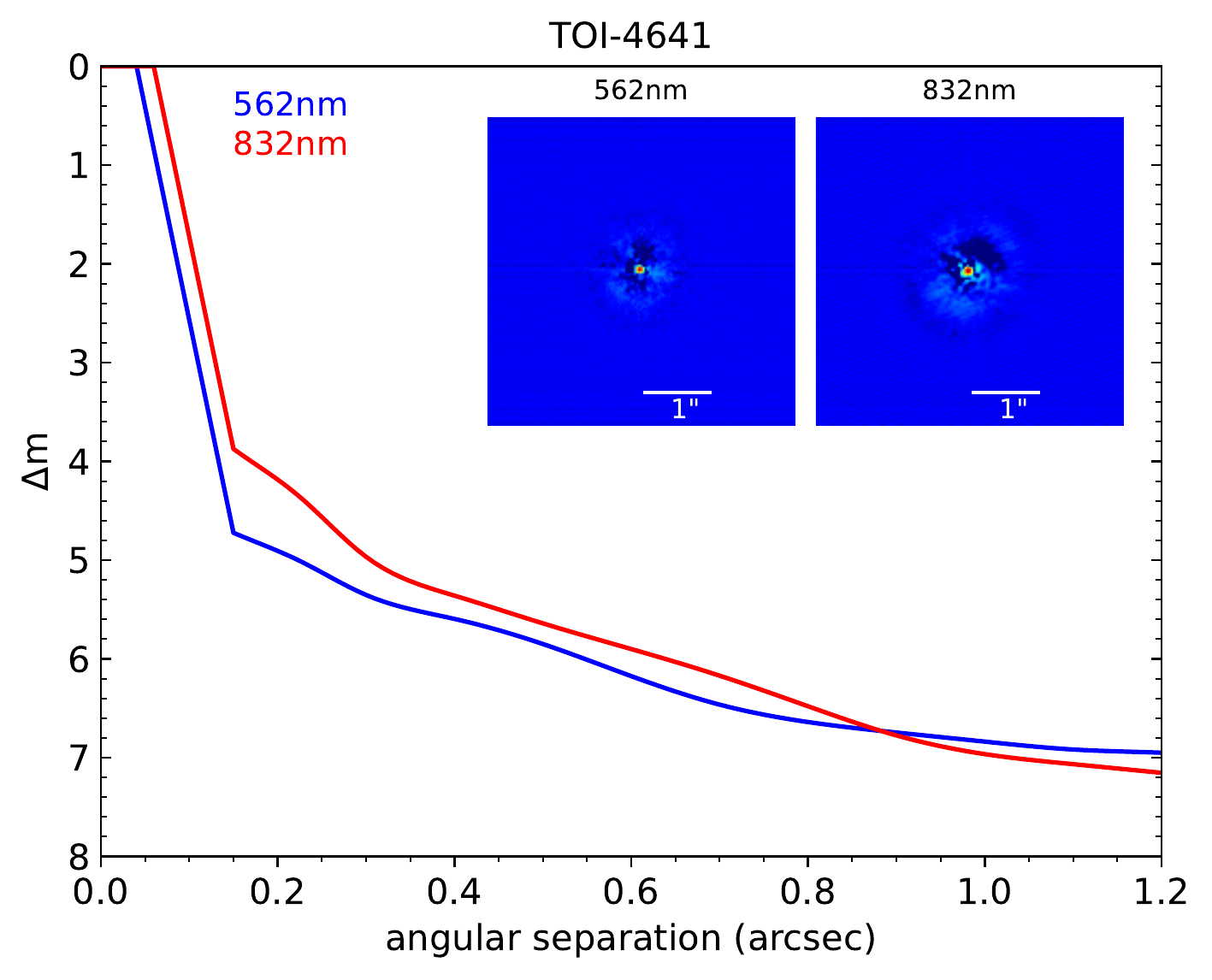} 
   \caption{Contrast curves based on NESSI speckle imaging of TOI-4641 in two filters centered at 562~nm (blue curve) and 832~nm (red curve).  Speckle imaging analysis was confined to the inner $1.2\arcsec$ of the full $4.6\times4.6\arcsec$ field-of-view.  Speckle reconstructed images centered on the star are inset at the top of the figure.  We find no companions to TOI-4641 in these high resolution imaging observations.}\label{fig:speckles}
\end{figure}

We observed \toi on UT February 5, 2023 using NESSI \citep{NESSI}, a speckle imager at the WIYN 3.5m telescope. The observations consisted of taking speckle sequence data in two filters with central wavelengths of 562 and 832nm.  We reduced these data using the standard speckle pipeline \citep{speckle2011} to obtain reconstructed images of the focal plane as well as a contrast curve centered on the target star and extending out to a radius of 1.2 arcseconds.  No secondary sources were detected surrounding \toi. The background limit plots are shown in Figure \ref{fig:speckles}.

\subsection{Spectroscopic Observations} \label{subsec:spec}

\subsubsection{Radial Velocity Observations} \label{subsec:TRES}
A single observation of \toi was obtained on UT November 23, 2021 with the NRES \citep{Siverd18} spectrograph located at the Las Cumbres Observatory \citep[LCOGT;][]{Brown13} node at McDonald Observatory, Texas, USA. Two successive 900 second observations were stacked and reduced via the BANZAI-NRES pipeline \citep{McCully22} to estimate a $v\sin I_\star$ of approximately 90 \kms, indicating that \toib was a favorable target for further investigation into the projected spin-orbit angle.

 Spectroscopic observations were obtained with the Tillinghast Reflector Echelle Spectrograph (TRES; \cite{gaborthesis}) as part of the TESS Follow-up Observing Program (TFOP) SubGroup 2 (SG2) Reconnaissance Spectroscopy program. TRES is an optical (3900-9100\,\AA) fiber-fed echelle spectrograph on the 1.5m Tillinghast Reflector at the Fred Lawrence Whipple Observatory (FLWO) in Arizona, USA. The spectrograph has a resolving power of R $\sim$ 44,000. 
 
 \toi was first observed on UT November 30, 2021 with an exposure time of 180 seconds leading to a signal-to-noise per resolution element (SNRe) of 113. The quick-look classification showed that the host star was rapidly rotating. Despite the rapid rotation which would make precise radial velocities very challenging, a second TRES spectrum (SNRe of 131) at opposite orbital quadrature was obtained on UT December 21, 2021 to check for a large velocity variation that would be indicative of a stellar companion.  The velocity offset between the two spectra was $\sim 275$ \ms\, based on the standard TRES pipeline which derives velocities using a single order of the spectrum centered on the Mg\,b features as described in \cite{buchhave2010}. An additional 13 TRES observations were obtained between UT December 10-31, 2022 with longer exposures than the original reconnaissance spectra in an attempt to gain high enough SNRe to detect an orbital solution. The average SNRe of the new observations was 195. 

In an attempt to get the best velocity precision, we used a least squares deconvolution technique \citep{zhou2016} based on the methods of \cite{donati1997} and \cite{colliercameron2010b} to extract radial velocities. Due to the rapid rotation of the host star, we were unable to detect an orbital signal of the planet. We also tried a multi-spectral order analysis of the available spectra.  Each spectrum was cross-correlated order-by-order against the highest SNRe observed spectrum to derive relative multi-order velocities. While we again were unable to detect a clear orbital signal, we were able to determine a 3$\sigma$ upper limit planetary mass of 3.87 $\mj$. The least squares deconvolution velocities are presented in Table~\ref{tab:rvs}, and were adopted for further analysis in Section~\ref{sec:global}. In addition, we also modeled the line profile determined from each spectrum to derive rotational and macroturbulent broadening velocities for the host star. The line profile is modeled as per \citet{1994AJ....107..742G}, with the macroturbulent broadening component described via a radial-tangential model. We find a best-fit rotational broadening velocity of \vsiniLSD\,\kms{}, and a macroturbulent velocity of \macroturb\,\kms for TOI-4641. 

\subsubsection{Transit Spectroscopic Observation} \label{subsec:DT}
As a planet transits across its host star's stellar disk, a portion of the stellar blue- and red-shifted light is blocked which produces a shift in the radial velocity measurements. This effect, known as the Rossiter-McLaughlin (RM; \cite{rossiter}, \cite{mclaughlin}) effect, allows us to measure the projected spin-orbit angle of a transiting system. 

To determine the projected spin-orbit angle, spectra were obtained during two transits on UT December 17, 2022 and UT January 8, 2023 using the TRES spectrograph. Spectra were acquired in the standard way by obtaining a set of three 300 second exposures surrounded on either side by Thorium-Argon calibration spectra. The three spectra were then combined using cosmic ray rejection and run through the standard TRES pipeline as described in Section \ref{subsec:TRES}. We observed a partial ingress on the night of UT December 17, 2022 collecting 18 sets of 3 spectra and a full transit on UT January 8, 2023 of 39 sets of 3 spectra with an average SNRe of 147 and 195, respectively. 

\begin{figure*}
     \centering
\begin{tabular}{ccc}
\textbf{2022-12-17} & \textbf{2023-01-08}  & \textbf{Joint modeling} \\
  \includegraphics[width=0.6\columnwidth]{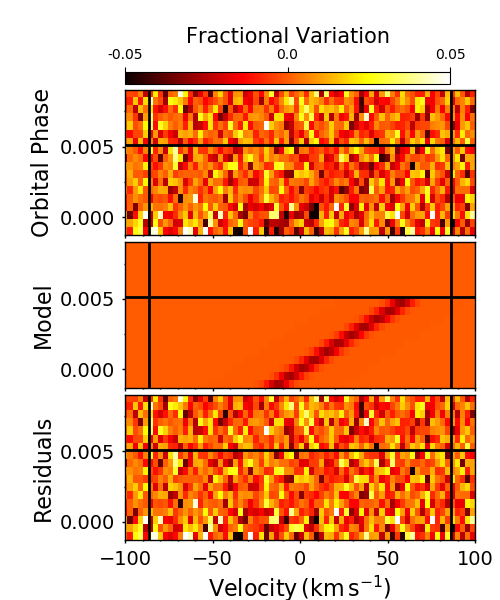} &
  \includegraphics[width=0.6\columnwidth]{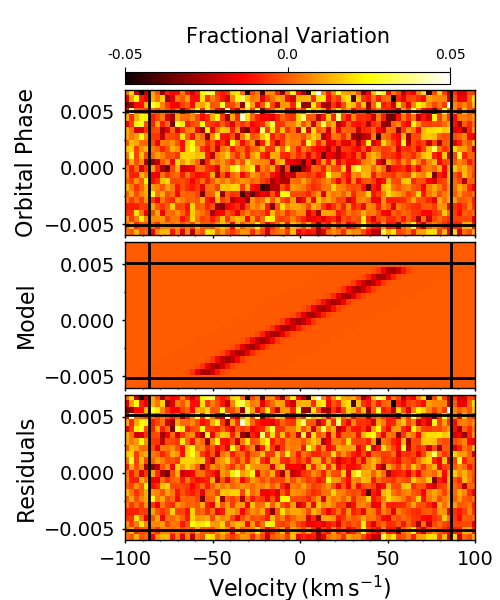} &
    \includegraphics[width=0.6\columnwidth]{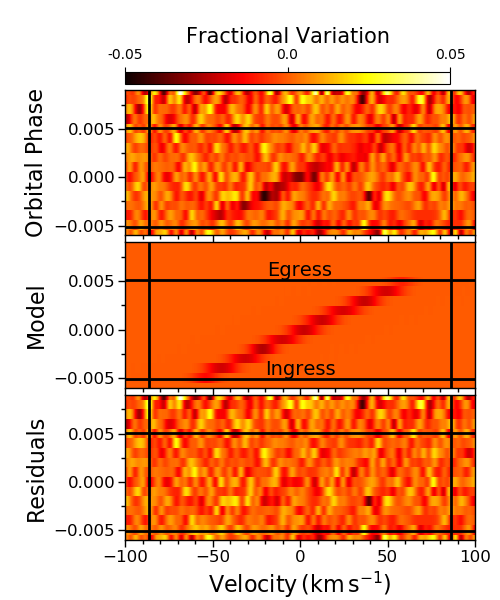} \\
\end{tabular}
  \caption{TRES Doppler spectroscopic results. The color scale represents the fractional variation in the spectral line profile. Each plot shows the planetary signal after the average rotational profile is subtracted (top), the best-fit model (middle), and the residuals after subtracting the planetary signal (bottom). \textbf{Left:} Doppler spectroscopy egress transit event on December 17, 2022. \textbf{Middle:} Doppler spectroscopy observations from the full transit event on January 8, 2023 \textbf{Right:} The combined Doppler spectroscopy result from the two TRES transit observations. The vertical lines are the \vsini\, boundaries, while the horizontal lines mark the ingress and egress timings.}\label{fig:DT}
\end{figure*}

To extract the planetary Doppler shadow, we derived the line broadening profiles for each observation in an analysis similar to that described above. We performed a least squares deconvolution between the observed spectra and a synthetic ATLAS9 stellar template \citep{Castelli:2004}, generated at the atmospheric parameters of the host star, with no rotational broadening incorporated. We then modeled the differences between each derived line profile and the median combined line profile measured over the transit night as part of our global modeling analysis as per \citet{2019AJ....157...31Z}. Briefly, for each observation, we calculated the integrated line profile of the portion of the star blocked by the planet. This incorporates the effects of local limb darkening, macroturbulent broadening, and rotation. This is modeled simultaneously with the system parameters as part of the global modeling process (Section~\ref{sec:global}). The line profile residuals for each transit observation, and for the combined observations, are shown in Figure~\ref{fig:DT}. The dark trail from bottom left to top right represents the shadow of the planet during the transit in the line profile residuals. 

%%%%% RVs %%%%%%%%%

\begin{table}
    \centering
    \caption{TRES Recon Radial Velocities}
    \label{tab:rvs}
    \begin{tabular}{rrr}
    \hline\hline
    \textbf{Time} & \textbf{Velocity} & \textbf{Uncertainty} \\
    \textbf{[\bjdtdb]} & \textbf{[\kms]} & \textbf{[\kms]} \\
    \hline
2459548.73116 & 24.12 & 0.26  \\
2459560.66830 &  24.35 & 0.22  \\
2459923.66941 & 24.34 & 0.29  \\
2459924.68547 & 24.58 & 0.31  \\
2459925.70482 & 24.53 & 0.21  \\
2459928.73172 & 24.31 & 0.27  \\
2459929.73860 & 24.19 & 0.32  \\
2459933.80715 & 24.56 & 0.31  \\
2459934.80076 & 24.49 & 0.28  \\
2459935.67590 & 24.58 & 0.30  \\
2459936.82834 & 24.53 & 0.22  \\
2459937.70062 & 24.44 & 0.27  \\
2459938.74970 & 24.52 & 0.30  \\
2459939.81605 & 24.51 & 0.26  \\
2459944.75598 & 24.34 & 0.30  \\
\hline
    \end{tabular}
\end{table}

%%%%%%%%%%%%%%%%%%%%%%%%%%%%%%%%%%%%%%%%%%%%%%%%%%%%%%%%%%%%%%%%%%%%%%%%%
\begin{table}
    \caption{Stellar parameters for \toi.
    \label{tab:star}}
    \centering
    \begin{tabular}{lrl}
    \hline\hline
\textbf{Stellar Parameters} & \textbf{Value} & \textbf{Source}\\
\hline
\multicolumn{3}{l}{\textbf{Catalog Information}}\\
TIC ID & 436873727 &  TESS TOI Catalog \\
TOI ID & 4641 & TESS TOI Catalog \\
\textit{Gaia} DR3 ID & 114340658009875072 & GAIA DR3\\
\textit{2MASS} ID & J02501388+2520010 & 2MASS\\
TYC ID & 1785-00801-1 & Tycho\\
%HIP ID & 13224 & \\
\multicolumn{3}{l}{\textbf{Coordinates and Proper Motion}} \\
Right Ascension (h:m:s)  & 02:50:13.90 (J2000) &  GAIA DR3 \\
Declination (d:m:s)    & 25:20:01.00 (J2000) &  GAIA DR3 \\
Parallax (mas)   & \parallax &  GAIA DR3 \\
$\mu$\textsubscript{R.A} (mas yr\textsuperscript{-1}) & \pmRA & GAIA DR3 \\
$\mu$\textsubscript{Dec.} (mas yr\textsuperscript{-1}) & \pmDEC &  GAIA DR3 \\
\multicolumn{3}{l}{\textbf{Magnitudes}} \\
$TESS$ (mag)         & $7.1323 \pm 0.0061$ & TESS TOI Catalog \\
$G$ (mag)         & 7.42334~$\pm$~0.00031 & GAIA DR3 \\
$B_{p}$ (mag)         & 7.60660~$\pm$~0.00093 & GAIA DR3 \\
$R_{p}$ (mag)         & 7.09227~$\pm$~0.00057 & GAIA DR3 \\
$B$ (mag)      & 7.895~$\pm$~0.027 & Tycho \\
$V$ (mag)         & 7.51~$\pm$~0.030 & Tycho \\
$J$ (mag)         & 6.746~$\pm$~0.020 & 2MASS \\
$H$ (mag)         & 6.639~$\pm$~0.023 & 2MASS \\
$K$ (mag)        & 6.579~$\pm$~0.029 & 2MASS\\
$WISE_{3.4\mu}$ (mag)     & 6.552~$\pm$~0.081 & WISE \\
$WISE_{4.6\mu}$ (mag)     & 6.527~$\pm$~0.024 & WISE \\
$WISE_{12\mu}$ (mag)     & 6.588~$\pm$~0.016 & WISE \\
$WISE_{22\mu}$ (mag)     & 6.426~$\pm$~0.085 & WISE \\
\hline
\end{tabular}
\begin{flushleft}
\footnotesize{\textbf{Note:} TESS TOI Primary Mission Catalog; \citep{Guerrero2021}, Tycho; \citep{tycho}; GAIA DR3; \citep{Gaiadr3}, 2MASS; \citep{2MASS}, WISE; \citep{allwise}}
\end{flushleft}
\end{table}

\subsubsection{Stellar Parameters from TRES spectra} \label{subsec:SPC}
We used the Stellar Parameter Classification (SPC; \citealt{buchhave2012}) tool to derive stellar parameters using the TRES spectra. SPC cross-correlates an observed spectrum against a library grid of synthetic spectra calculated using the \cite{kurucz1992} atmospheric models. A $\sim310$ \AA\, region of the spectrum surrounding the Mg b lines is used to derive effective temperature, \teff, surface gravity, \logg, rotational velocity, \vsini, and metallicity, \meh. Metallicity is derived using all available metal lines rather than just the Fe lines and therefore reported as \meh\, but is closely related to \feh\, values. 

%The line broadening parameter from SPC is designated \vsini\, but it does not correct for macroturbulence. 

\begin{figure}
    \centering
    \includegraphics[width=0.95\columnwidth]{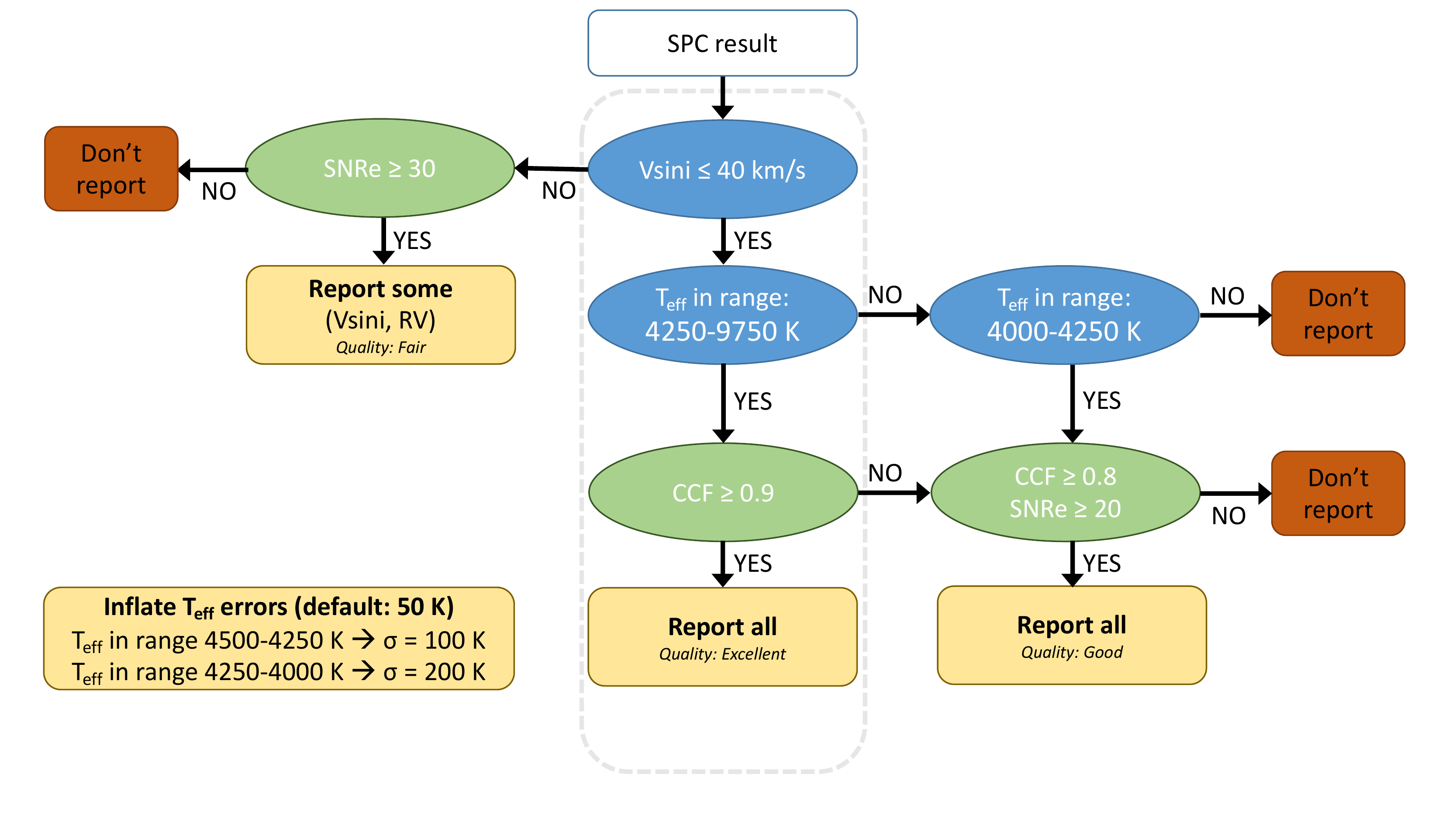}
    \caption{SPC quality flag decision tree assigns a quality flag - Excellent, Good, Fair, and Poor - based on the known limitations of SPC. The quality flag reliability of the result uses the stellar effective temperature, rotational velocity, SNRe, and the cross-correlation function (CCF) peak value as quality indicators.}\label{fig:SPC}
\end{figure}

We recently developed a quality flag (QF) metric - Excellent, Good, Fair, and Poor - for SPC results based on the known limitations of SPC. Each spectrum is run through an algorithm (shown in Figure \ref{fig:SPC}) to determine the reliability of the result using stellar effective temperature, rotational velocity, SNRe, and the cross-correlation function (CCF) peak value as quality indicators. As a service to the community, SPC stellar parameters for all TRES spectra of \tess targets are uploaded to the ExoFOP website when the QF is Excellent or Good. If the QF is determined to be Fair, as in the case of \toi due to the high stellar rotational velocity, the stellar parameters are not considered reliable and only the rotational velocity of the star is uploaded. Stellar parameters are not uploaded when a QF of Poor is determined.

SPC reports a \vsini\, of \vsiniSPC\, \kms\, for \toi\, but because SPC does not solve for macroturbulence and the Least Squares Deconvolution analysis does, we chose to use the Least Squares Deconvolution reported value as determined in Section~\ref{subsec:TRES} (\vsiniLSD\, \kms\,) as a prior for our global analysis. The SPC uncertainty reported is the floor error and should be inflated due to the star's rapid rotation.

%  of $-0.2 \pm\, 0.2$., in addition to a global fit with solar metallicity as a starting value. Both fits were consistent and 

\subsection{SED analysis} \label{sec:sed}
\begin{figure}
    \centering
    \includegraphics[width=0.46\textwidth]{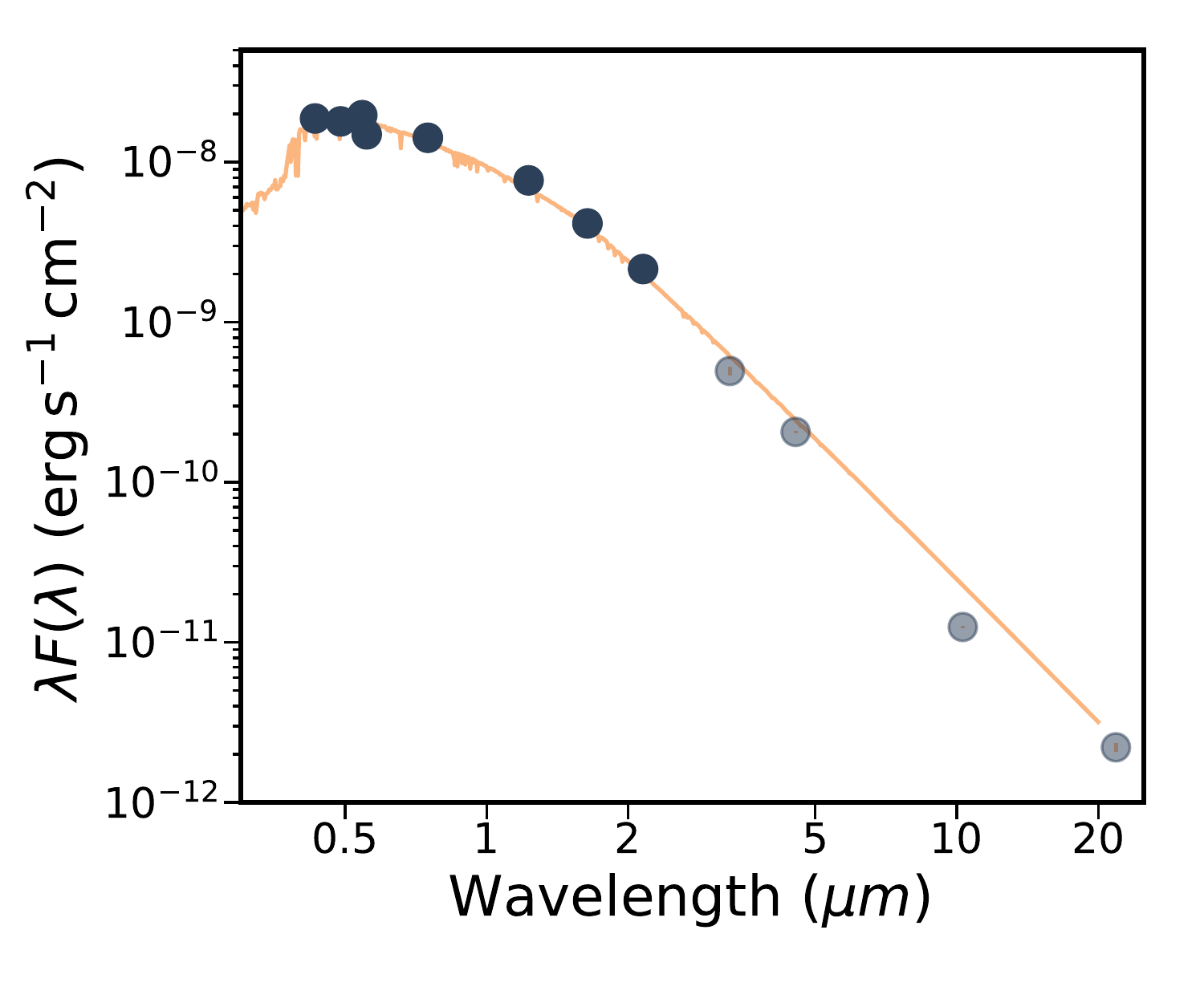}
    \caption{Spectral energy distribution of the target star TOI-4641. Magnitudes from \emph{Gaia} $G$, $B_p$, $R_p$, \emph{Tycho} $B$, $V$, 2MASS $J$, $H$, $Ks$ are included in the global modeling of the system and are shown in dark blue. Magnitudes from WISE $W1$, $W2$, $W3$, and $W4$ are shown in light grey but not included in the model. }\label{fig:sed}
\end{figure}

We used all available broadband photometry, including \emph{Hipparcos} $B$ and $V$ bands \citep{1997AA...323L..49P}, \emph{Gaia} DR3 $G$, $Bp$, $Rp$ \citep{2022arXiv220800211G}, 2MASS $J$, $H$, $K$ \citep{2006AJ....131.1163S}, and WISE $W_1$, $W_2$, $W_3$, $W_4$ bands \citep{WISE}, as well as \emph{Gaia} DR3 parallaxes to model the spectral energy distribution of \toi. The spectral energy distribution is modeled simultaneously with the photometric and spectroscopic observations of the system, such that the stellar properties derived are jointly constrained by the transit and the photometric properties of the star. At each iteration of the global model (Section~\ref{sec:global}), we compute the interpolated isochrone magnitudes for each tested stellar mass, age, and metallicity. We then compute the log likelihood between the isochrone magnitudes and the observed values and associated uncertainties for each band. We adopt the MIST isochrones \citep{2016ApJS..222....8D}, interpolated via the \textsc{minimint} package \citep{2020zndo...4002972K}, for our stellar models. The best fitting model, as per Table~\ref{tab:star}, is shown in Figure~\ref{fig:sed}. 

\section{Global Modeling} \label{sec:global}

To determine the stellar and planetary parameters of the system holistically, we performed a joint analysis of all available photometric, spectroscopic, and catalog observations. This included photometric transit observations from TESS and \textit{Tierras} (Section \ref{subsec:TESS} and \ref{subsec:Tierras}), two spectroscopic transit observations from TRES (Section \ref{subsec:DT}), the TRES out-of-transit radial velocities (Section \ref{subsec:TRES}), and available catalog observations (Section \ref{sec:sed}). Free parameters largely describing the transit include the orbital period $P$, reference time of transit center $T_0$, planet-to-star radius ratio $R_p/R_\star$, line-of-sight inclination $i$, and orbital eccentricity parameters $\sqrt{e}\cos \omega$ and $\sqrt{e} \sin \omega$. In addition, the radial velocity orbit is modeled by including the free parameter describing the mass of the planetary companion $M_p$. The spectroscopic transit is modeled as per \citet{2019AJ....157...31Z}, with free parameters including the projected spin-orbit angle $\lambda$, rotational broadening $v\sin I_\star$, and macroturbulent broadening velocity. Simultaneous with the transit models, we also interpolate the stellar isochrones as per Section~\ref{sec:sed}. At each step, we model the spectral energy distribution to constrain the stellar parameters. Free parameters describing the stellar isochrone and spectral energy distribution modeling include stellar mass $M_\star$, age, metallicity [m/H], and parallax. 

A number of parameters are constrained by informed priors in the global modeling. Parallax is tightly constrained by a Gaussian prior about its \emph{Gaia} DR3 value and associated uncertainties. Rotational and macroturbulent broadening velocities are constrained by Gaussian priors about their spectroscopically determined values. Stellar metallicity is constrained by a Gaussian prior about the SPC-determined value. Uniform priors about reasonable physical parameter spaces are adopted for all other free parameters. The adopted priors are noted in Table~\ref{tab:bestfit}. Gaussian priors are noted as $\mathcal{G}(\mu,\sigma)$, while uniform priors and their adopted ranges are noted by $\mathcal{U}(\mathrm{min},\mathrm{max})$. 

The photometric transits are modeled as per \citet{mandelAgol2002} via the \textsc{batman} package \citep{2015PASP..127.1161K}. Limb darkening parameters are interpolated and fixed to their values as per \citet{claret2011}, \citet{2017A&A...600A..30C}, and \citet{eastman2013}. The spectroscopic transit is modeled as per Section~\ref{subsec:DT}, via a disk integration of the portion of the stellar surface occulted by the planet, incorporating the effect of local macroturbulence and rotational broadening. 

The best fit parameters are presented in Table~\ref{tab:star} and Table~\ref{tab:bestfit}. A number of additional parameters are derived from the posterior chains and reported in Table~\ref{tab:bestfit} for completeness. These are marked as `inferred' in the table. Stellar parameters for luminosity, effective temperature, surface gravity, and age are re-derived from MIST isochrones interpolations for each given link in the MCMC chain subsequent to the global modeling. Planet properties, including radius, orbital semi-major axis, eccentricity, and transit impact parameter are subsequently derived from the posteriors. In addition to the global model, we also propagate the posteriors and follow \citet{2020AJ....159...81M} to derive the 3D spin-orbit obliquity of the system. We make use of the rotational period (see Section~\ref{sec:variability}), line broadening, and projected spin-orbit angle to derive a 3D orbital obliquity of $2.4\pm1.3^\circ$. 

\begin{table*}
    \caption{Best-fit Stellar and Planetary Properties for \toi   \label{tab:bestfit}}
    \centering
    \begin{tabular}{lllr}
    \hline\hline
    \textbf{Parameters} &
    \textbf{Description (Units)} & \textbf{Prior Values} &
    \textbf{Best Fit} \\
    \hline
\multicolumn{4}{l}{\textbf{Stellar Parameters:}} \\
$M_{\star}$ & Stellar Mass ($M_\odot$) & $\mathcal{U}(1.2,3.5)$ & \mstar \\
$R_{\star}$ & Stellar Radius ($R_{\odot}$) & $\mathcal{U}(1.2,3.5)$ & \rstar  \\
$L_{\star}$ & Stellar Luminosity ($L_{\odot}$) & Inferred & \lstar \\
$T_{\rm eff}$ & Effective Temperature (K)  & Inferred & \tefffit \\
$\log g$ & Surface Gravity (cgs)   & Inferred & \loggfit \\
$\mathrm{[m/H]}$ & Metallicity (dex) & $\mathcal{G}(-0.2,0.2)$ & \metfit \\
$v \sin I_\star$ & Projected Rotational Velocity (\kms)  & $\mathcal{G}(86.3,1.0)$ & \vsinifit   \\
$v_\mathrm{macro}$ & Macroturbulent Velocity (\kms) & $\mathcal{G}(5.5,7.7)$ & \macroturb \\
Parallax & Parallax (mas) & $\mathcal{G}(11.409,0.026)$ & $11.409 \pm\, 0.025$ \\
Age & Age (Gyr) & Inferred & \age \\
Distance & Distance (pc) & Inferred & \dist\\
$I_\star$ & Stellar Inclination (deg) & Derived & $>83$ $(3\sigma)$ \\
$u_\mathrm{1,TESS}$ & Limb darkening coefficient & Fixed & 0.197 \\ %SEARCH FOR IN CONFIG file
$u_\mathrm{2,TESS}$ & Limb darkening coefficient & Fixed & 0.317 \\
$u_\mathrm{1,Tierras}$ & Limb darkening coefficient & Fixed & 0.209 \\
$u_\mathrm{2,Tierras}$ & Limb darkening coefficient & Fixed & 0.314 \\
\hline
    \multicolumn{4}{l}{\textbf{Planetary Parameters:}} \\
$P$ & Orbital Period (days) & $\mathcal{U}(22.09,22.10)$ & \per \\
$T_{o}$ & Epoch (BJD) & $\mathcal{U}(245910.80,245910.85)$ & \midt \\ 
$M_{p}$ & Planet Mass (\mj) & $\mathcal{U}(0,105)$ & $< 3.87 \, (3\sigma)$ \\
$R_{p}$ & Planet Radius (\rj) & Inferred & \plrad \\
$R_{p}/R_{\star}$ & Radius of planet to star ratio  &  $\mathcal{U}(0.01,0.1)$ & \rprs \\
$a/R_{\star}$ & Semi-major axis to star radius ratio  & Inferred & \aor \\
$a$ & Semi-major axis (AU) & Inferred & \semimaj \\
$e$ & eccentricity  & Inferred & $< 0.074\,(3\sigma)$  \\
$\sqrt{e} \cos \omega$ & eccentricity parameter & $\mathcal{U}(-0.2,0.2)$ & $0.01_{-0.14}^{+0.13}$ \\
$\sqrt{e} \sin \omega$ & eccentricity parameter & $\mathcal{U}(-0.2,0.2)$ & $0.03_{-0.13}^{+0.12}$ \\
$\gamma$ & Radial velocity offset (\kms) & $\mathcal{U}(24,25)$ & $24.400_{-0.072}^{+0.077}$\\
$T_{14}$ & Transit duration (days) & Inferred & \tdur \\
$i$ & Orbital inclination (deg) & $\mathcal{U}(80,110)$ & \incl \\
$b$ & Impact parameter & Inferred & \bimp \\
$\lambda$ & Projected spin-orbit angle (deg) &  $\mathcal{U}(-185,185)$ & \lam \\
$\psi$ & 3D orbital obliquity (deg) & Derived & $2.4\pm1.3$ \\    
    \hline
    \end{tabular}
\begin{center}
\footnotesize{\textbf{Note:} Gaussian priors are listed as $\mathcal{G}$(median,width) and uniform priors are listed as $\mathcal{U}$(lower bound, upper bound). Inferred parameters are calculated from the posterior distribution. TRES spectra were used to derive a metallicity prior using the Stellar Parameter Classification analysis \citep{buchhave2012} and the project rotational velocity and macroturbulent velocity using a least squares deconvolution analysis \citep{zhou2016}. Parallax was obtained from the \gaia DR3 release \citep{Gaiadr3}.}    
\end{center}
\end{table*}

\section{Stellar variability} \label{sec:variability}

TOI-4641 exhibits short timescale stellar variability at the 1\,mmag level during all five sectors of \emph{TESS} observations and is also noticed in the ground-based observations. The variability is likely consistent with changing rotational spot modulation on the stellar surface.   

\begin{figure}
    \centering
  \includegraphics[width=0.95\columnwidth]{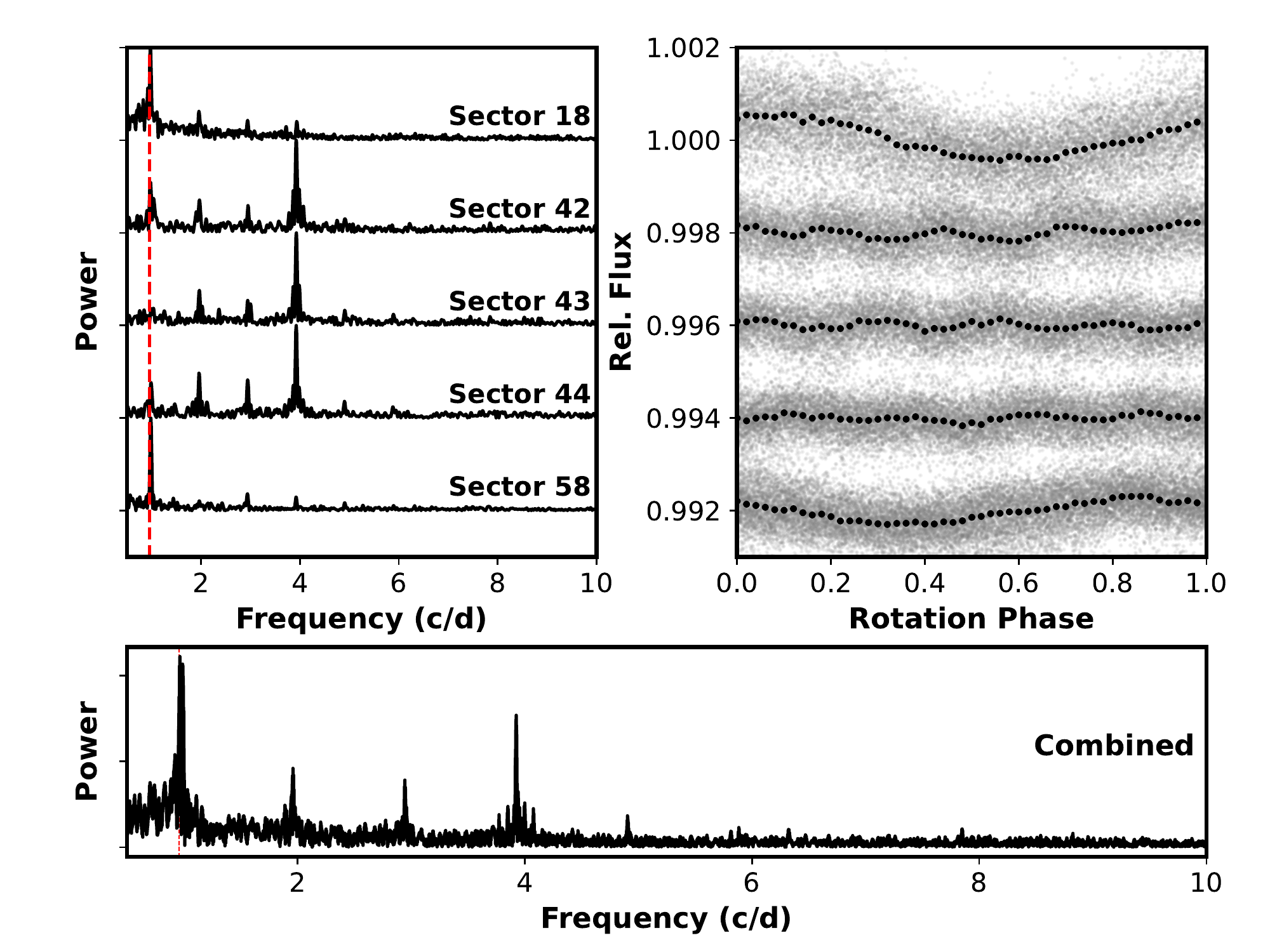}
    \caption{TOI-4641 exhibits 1 day time scale variability over all observed \emph{TESS} sectors. The per sector frequency spectra are shown in the \textbf{top left} panel. The frequency spectrum of the entire \emph{TESS} dataset is shown on the \textbf{bottom}. The per sector light curves, folded to the $0.96$ cycles/day frequency, are shown on the \textbf{top right} panel. The 0.96 cycle per day peak is consistent with the expected rotation period of the star as per the spectroscopic rotational line broadening velocity. We interpret the variability to be spot-induced modulations, with the spots likely changing in configuration over multiple months.}
    \label{fig:pulsations}
\end{figure}

The top left panel of Figure~\ref{fig:pulsations} shows the sector-by-sector frequency power spectrum of TOI-4641. Sectors 18 and 58 exhibit variability at the $0.96\pm0.02$ cycles/day frequency, while the $3.93\pm0.02$ cycles/day peak is strongest for Sectors 42, 43, and 44. The variability does not exhibit equal period spacing as is expected for $\gamma$ Dor variables, and the frequency is too low for $\delta$ Scuti pulsations. We suggest the variability is most consistent with rotational modulation. The rotational broadening velocity $v \sin I_\star = 86.3\pm1.0 \, \mathrm{km\,s}^{-1}$ propagates to an expected rotation period of $1.06\pm0.04$ days assuming $\sin I_\star = 1$, corresponding to $0.94\pm0.04$ cycles/day, which is consistent with the peak variability frequency in the light curve in Sectors 18 and 58. The 1 cycle/day periodicity is the second most dominant peak in Sectors 42, 43, and 44, with the 4 cycles/day peak likely due to the specific spot configuration during this timeframe. Low-amplitude spot variability is seen in early-type stars in the \emph{Kepler} sample. \citet{2020MNRAS.498.2456S} notes that 10-30\% of A and B stars exhibit spot-induced variability, despite not showing spectroscopic signatures of chemical peculiarity. Using all \tess sectors of data, we derive a $3\sigma$ lower limit on the stellar inclination $I_\star$ of 83 $\degr$.

\begin{figure}
    \centering    \includegraphics[width=0.95\columnwidth]{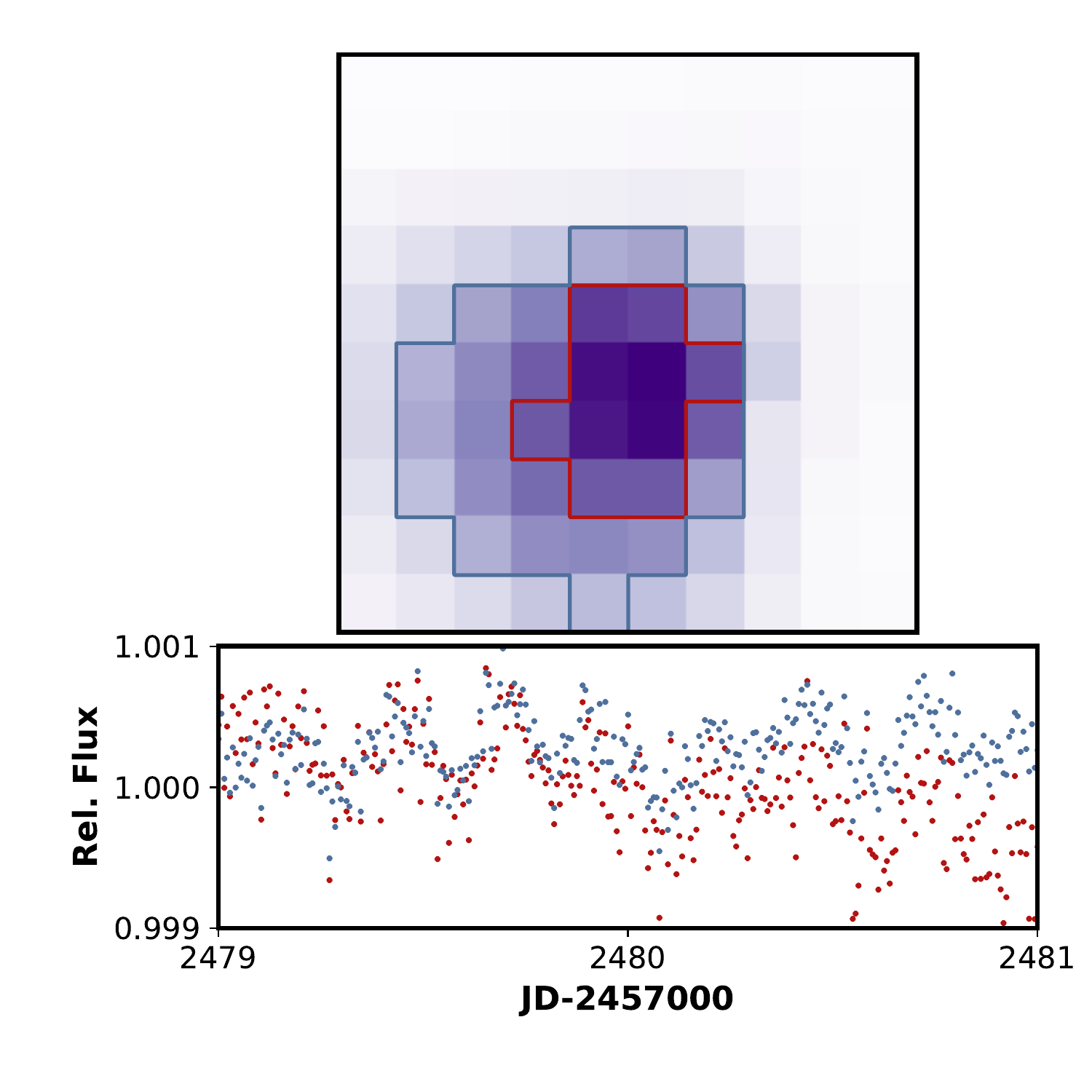}    
    \caption{Photometry, using two different apertures, showing that the observed variability is likely on target and not induced by a nearby background source. The \textbf{top} panel shows a narrow 90-percentile (red) and a broad 50-percentile (blue) aperture. The \textbf{bottom} panel shows a 2-day example section of their respective light curves showing no difference in the frequency and amplitude of the observed light curve as a function of aperture size. }
    \label{fig:pulsation_apertures}
\end{figure}

To check if the variability is on target, we used the \textsc{lightkurve} code \citep{lightkurve} to extract light curves from multiple target and background apertures, and found no changes in the amplitude and shape of the variability. Figure~\ref{fig:pulsation_apertures} shows a set of example target apertures and resulting light curves. The target is in a sparse field and there are no stars within a 1' field that are bright enough to cause the 1\,mmag variability seen on target after dilution. We also note that the rotational modulation was similarly identified in the SPOC transit search of the combined \toi light curves. These `Threshold Crossing Events' (TCEs) attributed to the rotational modulation were subjected to the same diagnostic tests as would any TCE triggered by a transiting planet. The difference image centroid offsets \citep{Twicken2018} showed that the source of the modulation was consistent with \toi at the 3$\sigma$ level and inconsistent at that level with all other TIC v8.2 objects.

\section{Discussion} \label{sec:discussion}

\toi is a warm Jupiter in a 22 day orbit around a bright (V=7.5), rapidly rotating (\vsinifit \kms) F-star. The orbit is nearly circular with eccentricity constrained at the 3$\sigma$ level to less than 0.074. This target was observed in five \tess sectors and was photometrically followed up by the \textit{Tierras} ground-based Observatory to rule out the false positive scenario of a nearby eclipsing binary contaminating the aperture. Additionally, we obtained high-resolution images from the Speckle imager on the WIYN 3.5m telescope and detected no secondary sources out to a radius of 1.2 arcseconds. TRES spectra allowed us to detect a 3-$\sigma$ upper limit planetary mass of 3.87 $\mj$. We also obtained two nights of TRES in-transit spectroscopic data to measure the projected spin-orbit angle of \lam$\degr$.

\begin{figure}
    \centering    \includegraphics[width=0.95\columnwidth]{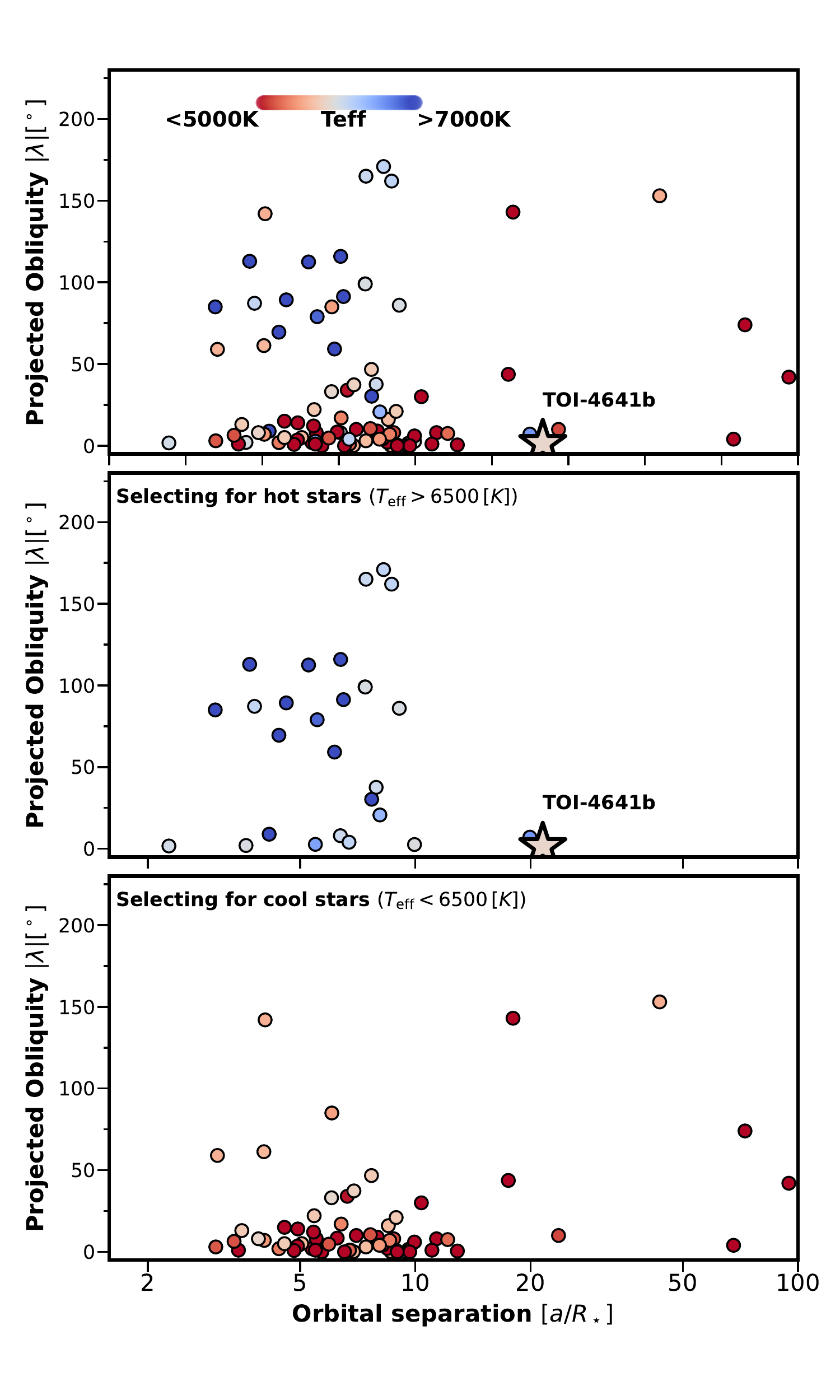}
    \caption{TOI-4641b amongst giant planets with obliquities measured. \textbf{Top} Obliquity as a function of $a/R_{\star}$. Values adopted from \citet{albrecht2022} including only planets with $R_{p} > 8\, R_{e}$ to represent the Jovian population. The colors represent the effective temperature of the host star. \toib is represented with a star. \textbf{Middle} Selecting only for planets about stars $T_\mathrm{eff} > 6500\,K$. TOI-4641 is one of the longest period planets characterized about an early-type star. \textbf{Bottom} The distribution for planets about cool stars $T_\mathrm{eff} < 6500\,K$.}
    \label{fig:obliqars_temp_albrecht}
\end{figure}

TOI-4641b is amongst the longest period planets to be thoroughly characterized about a hot rapidly rotating star (Figure~\ref{fig:obliqars_temp_albrecht}). Long-period planets provide tests for mechanisms that induce primordial misalignment in planetary systems. At such orbital distances, star-planet tidal interactions are too weak to modify the orbital obliquity.

Chaotic accretion is one proposed method of primordial misalignments where neighboring protostar material interacts or arrives at different times during the accretion process potentially causing the protostar disk to tilt with respect to the star \citep{bate2010, thies2011, fielding2015, bate2018, kuffmeier2021}. Another proposed mechanism for misalignment is magnetic warping. In young stars, in particular, the twisting of magnetic field lines between the ionized disk and the differential rotation of the young star can cause misalignment \citep{foucart2011, lai2012}. In this scenario, the misalignment torque must overcome the realignment torque from accretion, magnetic braking, disk winds and viscosity. Stellar or planetary companions can also cause misalignment during the primordial phase of formation \citep{borderies1984, lubow2000, batygin2012, matsakos2017}. \citet{2012ApJ...758L...6R} proposed that internal gravity waves excited at the radiative-convective boundary of early-type stars can induce their surface layers to change in spin direction. Changes to the spin axes of early-type stars will also lead to an apparent spin-orbit misalignment, potentially contributing to the temperature-obliquity gradient seen in the hot Jupiter population. 

Critically, most of these proposed mechanisms do not have a strong dependence on the host star properties beyond planet-star tidal interactions. We should not observe strong differences in the obliquity distributions of longer-period Jovian planets as a function of stellar mass. Should internal gravity waves \citep{2012ApJ...758L...6R} play a major role in shaping the spin axes of early-type stars, no orbital distance trends should be observed in that population. 

Testing these predictions motivate full characterizations of planets in long-period orbits about early-type stars. TOI-4641b is the second such Jovian-sized planet around a rapidly rotating early-type star, preceded only by Kepler-448b \citep{2015A&A...579A..55B,2017AJ....154..137J}. Both planets have been found in well-aligned geometries, whereas early-type stars with closer orbiting giant planets tend to show a broad range of obliquities including nearly-polar and retrograde orbits.

\section*{Acknowledgments}

%USQ
We respectfully acknowledge the traditional custodians of the lands on which we conducted this research and throughout Australia. We recognize their continued cultural and spiritual connection to the land, waterways, cosmos and community. We pay our deepest respects to all Elders, present and emerging people of the Giabal, Jarowair and Kambuwal nations, upon whose lands this research was conducted.
%Scholarship
CH thanks the support of the ARC DECRA program DE200101840.
GZ thanks the support of the ARC DECRA program DE210101893.
% Tierras
JGM acknowledges support by the Heising Simons Foundation through a 51 Pegasi B Fellowship, and by the Pappalardo family through the MIT Pappalardo Fellowship in Physics. JGM and DC thank the staff at the F. L. Whipple Observatory for their assistance in the refurbishment and maintenance of the 1.3-m telescope. Tierras is supported by grants from the John Templeton Foundation and the Harvard Origins of Life Initiative. The opinions expressed in this publication are those of the authors and do not necessarily reflect the views of the John Templeton Foundation.
SJM was supported by the Australian Research Council (ARC) through Future Fellowship FT210100485.
% TESS
Funding for the TESS mission is provided by NASA's Science Mission Directorate. We acknowledge the use of public TESS data from pipelines at the TESS Science Office and at the TESS Science Processing Operations Center. This research has made use of the Exoplanet Follow-up Observation Program website, which is operated by the California Institute of Technology, under contract with the National Aeronautics and Space Administration under the Exoplanet Exploration Program. Resources supporting this work were provided by the NASA High-End Computing (HEC) Program through the NASA Advanced Supercomputing (NAS) Division at Ames Research Center for the production of the SPOC data products. This paper includes data collected by the TESS mission that are publicly available from the Mikulski Archive for Space Telescopes (MAST). This work has made use of data from the European Space Agency (ESA) mission
{\it Gaia} (\url{https://www.cosmos.esa.int/gaia}), processed by the {\it Gaia}
Data Processing and Analysis Consortium (DPAC,
\url{https://www.cosmos.esa.int/web/gaia/dpac/consortium}). Funding for the DPAC
has been provided by national institutions, in particular the institutions
participating in the {\it Gaia} Multilateral Agreement.

\section*{Data Availability}
The data underlying this article will be shared on reasonable request to the corresponding author.



\bibliographystyle{mnras}
\bibliography{ref} % if your bibtex file is called example.bib

% \begin{thebibliography}{99}
% \bibitem[\protect\citeauthoryear{Author}{2012}]{Author2012}
% Author A.~N., 2013, Journal of Improbable Astronomy, 1, 1
% \bibitem[\protect\citeauthoryear{Others}{2013}]{Others2013}
% Others S., 2012, Journal of Interesting Stuff, 17, 198
% \end{thebibliography}

%%%%%%%%%%%%%%%%%%%%%%%%%%%%%%%%%%%%%%%%%%%%%%%%%%
% Affiliations \newline
% $^{1}$University of Southern Queensland, Centre for Astrophysics, West Street, Toowoomba, QLD 4350 Australia\\

%%%%%%%%%%%%%%%%%%%%%%%%%%%%%%%%%%%%%%%%%%%%%%%%%%

% Don't change these lines
\bsp	% typesetting comment
\label{lastpage}
\end{document}

%% file: authors.tex
\author[Bieryla, et al.]{\parbox{\textwidth}
{Allyson Bieryla\orcid{0000-0001-6637-5401},$^{1,2}$\thanks{E-mail: \texttt{abieryla@cfa.harvard.edu}}
George Zhou\orcid{0000-0002-4891-3517},$^{2}$
Juliana Garc\'ia-Mej\'ia\orcid{0000-0003-1361-985X},$^{1,3,4}$
Tyler R. Fairnington\orcid{0000-0002-0692-7822},$^{1}$
David W. Latham\orcid{0000-0001-9911-7388},$^{1}$
Brad Carter\orcid{0000-0003-0035-8769},$^{2}$
Jiayin Dong\orcid{0000-0002-3610-6953},$^{5,6}$
Chelsea X. Huang\orcid{0000-0003-0918-7484},$^{2}$
Simon J. Murphy\orcid{0000-0002-5648-3107},$^{2}$
Avi Shporer\orcid{0000-0002-1836-3120},$^{4}$
Karen A.\ Collins\orcid{0000-0001-6588-9574},$^{1}$
Samuel N. Quinn\orcid{0000-0002-8964-8377},$^{1}$
Mark~E.~Everett\orcid{0000-0002-0885-7215},$^{7}$
Lars A. Buchhave\orcid{0000-0003-1605-5666},$^{8}$
Ren\'{e} Tronsgaard\orcid{0000-0003-1001-0707},$^{8}$
David~Charbonneau\orcid{0000-0002-9003-484X},$^{1}$
Marshall C. Johnson\orcid{0000-0002-5099-8185},$^{9}$
Gilbert A. Esquerdo\orcid{0000-0002-9789-5474},$^{1}$
Michael Calkins,$^{1}$
Perry Berlind,$^{1}$
Jon~M.~Jenkins\orcid{0000-0002-4715-9460},$^{10}$
George~R.~Ricker\orcid{0000-0003-2058-6662},$^{4}$
Sara~Seager\orcid{0000-0002-6892-6948},$^{4,11,12}$
Joshua~N.~Winn\orcid{0000-0002-4265-047X},$^{13}$
Thomas~Barclay\orcid{0000-0001-7139-2724},$^{14}$
Ismael~Mireles\orcid{0000-0002-4510-2268},$^{15}$
Martin~Paegert\orcid{0000-0001-8120-7457},$^{1}$
and Joseph~D.~Twicken\orcid{0000-0002-6778-7552}$^{16}$
}
\\
\\
$^{1}$\CfA\\
$^{2}$\USQ \\
$^{3}$51 Pegasi b Fellow \\
$^{4}$\MITKavli \\
$^{5}$Flatiron Research Fellow \\
$^{6}$\FlatironCCA \\
$^{7}$NSF’s National Optical-Infrared Astronomy Research Laboratory, 950 N. Cherry Ave., Tucson, AZ 85719, USA \\
$^{8}$\DTU \\
$^{9}$Department of Astronomy, The Ohio State University, 4055 McPherson Laboratory, 140 West 18$^{\mathrm{th}}$ Ave., Columbus, OH 43210 USA \\
$^{10}$NASA Ames Research Center, Moffett Field, CA 94035, USA \\
$^{11}$Department of Earth, Atmospheric and Planetary Sciences, Massachusetts Institute of Technology, 77 Massachusetts Ave, Cambridge, MA 02139, USA \\
$^{12}$Department of Aeronautics and Astronautics, Massachusetts Institute of Technology, 77 Massachusetts Avenue, Cambridge, MA 02139, USA \\
$^{13}$Department of Astrophysical Sciences, Princeton University, 4 Ivy Lane, Princeton, NJ 08544, USA \\
$^{14}$NASA Goddard Space Flight Center, 8800 Greenbelt Road, Greenbelt, MD 20771, USA\\
$^{15}$Department of Physics and Astronomy, University of New Mexico, 210 Yale Blvd NE, Albuquerque, NM 87106, USA \\
$^{16}$SETI Institute, Mountain View, CA 94043, USA \\
}

%% file: main.bbl
\begin{thebibliography}{}
\makeatletter
\relax
\def\mn@urlcharsother{\let\do\@makeother \do\$\do\&\do\#\do\^\do\_\do\%\do\~}
\def\mn@doi{\begingroup\mn@urlcharsother \@ifnextchar [ {\mn@doi@} {\mn@doi@[]}}
\def\mn@doi@[#1]#2{\def\@tempa{#1}\ifx\@tempa\@empty \href {http://dx.doi.org/#2} {doi:#2}\else \href {http://dx.doi.org/#2} {#1}\fi \endgroup}
\def\mn@eprint#1#2{\mn@eprint@#1:#2::\@nil}
\def\mn@eprint@arXiv#1{\href {http://arxiv.org/abs/#1} {{\tt arXiv:#1}}}
\def\mn@eprint@dblp#1{\href {http://dblp.uni-trier.de/rec/bibtex/#1.xml} {dblp:#1}}
\def\mn@eprint@#1:#2:#3:#4\@nil{\def\@tempa {#1}\def\@tempb {#2}\def\@tempc {#3}\ifx \@tempc \@empty \let \@tempc \@tempb \let \@tempb \@tempa \fi \ifx \@tempb \@empty \def\@tempb {arXiv}\fi \@ifundefined {mn@eprint@\@tempb}{\@tempb:\@tempc}{\expandafter \expandafter \csname mn@eprint@\@tempb\endcsname \expandafter{\@tempc}}}

\bibitem[\protect\citeauthoryear{{Albrecht} et~al.,}{{Albrecht} et~al.}{2012}]{albrecht2012b}
{Albrecht} S.,  et~al., 2012, \mn@doi [\apj] {10.1088/0004-637X/757/1/18}, \href {https://ui.adsabs.harvard.edu/abs/2012ApJ...757...18A} {757, 18}

\bibitem[\protect\citeauthoryear{{Albrecht}, {Dawson}  \& {Winn}}{{Albrecht} et~al.}{2022}]{albrecht2022}
{Albrecht} S.~H.,  {Dawson} R.~I.,   {Winn} J.~N.,  2022, \mn@doi [\pasp] {10.1088/1538-3873/ac6c09}, \href {https://ui.adsabs.harvard.edu/abs/2022PASP..134h2001A} {134, 082001}

\bibitem[\protect\citeauthoryear{{Bate}}{{Bate}}{2018}]{bate2018}
{Bate} M.~R.,  2018, \mn@doi [\mnras] {10.1093/mnras/sty169}, \href {https://ui.adsabs.harvard.edu/abs/2018MNRAS.475.5618B} {475, 5618}

\bibitem[\protect\citeauthoryear{{Bate}, {Lodato}  \& {Pringle}}{{Bate} et~al.}{2010}]{bate2010}
{Bate} M.~R.,  {Lodato} G.,   {Pringle} J.~E.,  2010, \mn@doi [\mnras] {10.1111/j.1365-2966.2009.15773.x}, \href {https://ui.adsabs.harvard.edu/abs/2010MNRAS.401.1505B} {401, 1505}

\bibitem[\protect\citeauthoryear{{Batygin}}{{Batygin}}{2012}]{batygin2012}
{Batygin} K.,  2012, \mn@doi [\nat] {10.1038/nature11560}, \href {https://ui.adsabs.harvard.edu/abs/2012Natur.491..418B} {491, 418}

\bibitem[\protect\citeauthoryear{{Borderies}, {Goldreich}  \& {Tremaine}}{{Borderies} et~al.}{1984}]{borderies1984}
{Borderies} N.,  {Goldreich} P.,   {Tremaine} S.,  1984, \mn@doi [\apj] {10.1086/162423}, \href {https://ui.adsabs.harvard.edu/abs/1984ApJ...284..429B} {284, 429}

\bibitem[\protect\citeauthoryear{{Bourrier} et~al.,}{{Bourrier} et~al.}{2015}]{2015A&A...579A..55B}
{Bourrier} V.,  et~al., 2015, \mn@doi [\aap] {10.1051/0004-6361/201525750}, \href {https://ui.adsabs.harvard.edu/abs/2015A&A...579A..55B} {579, A55}

\bibitem[\protect\citeauthoryear{{Bourrier}, {Cegla}, {Lovis}  \& {Wyttenbach}}{{Bourrier} et~al.}{2017}]{WASP8_2017}
{Bourrier} V.,  {Cegla} H.~M.,  {Lovis} C.,   {Wyttenbach} A.,  2017, \mn@doi [\aap] {10.1051/0004-6361/201629973}, \href {https://ui.adsabs.harvard.edu/abs/2017A&A...599A..33B} {599, A33}

\bibitem[\protect\citeauthoryear{{Brown} et~al.,}{{Brown} et~al.}{2013}]{Brown13}
{Brown} T.~M.,  et~al., 2013, \mn@doi [\pasp] {10.1086/673168}, \href {https://ui.adsabs.harvard.edu/abs/2013PASP..125.1031B} {125, 1031}

\bibitem[\protect\citeauthoryear{{Buchhave} et~al.,}{{Buchhave} et~al.}{2010}]{buchhave2010}
{Buchhave} L.~A.,  et~al., 2010, \mn@doi [\apj] {10.1088/0004-637X/720/2/1118}, \href {https://ui.adsabs.harvard.edu/abs/2010ApJ...720.1118B} {720, 1118}

\bibitem[\protect\citeauthoryear{Buchhave et~al.,}{Buchhave et~al.}{2012}]{buchhave2012}
Buchhave L.~A.,  et~al., 2012, Nature, 486, 375

\bibitem[\protect\citeauthoryear{{Castelli} \& {Kurucz}}{{Castelli} \& {Kurucz}}{2004}]{Castelli:2004}
{Castelli} F.,  {Kurucz} R.~L.,  2004, ArXiv Astrophysics e-prints, \href {http://adsabs.harvard.edu/abs/2004astro.ph..5087C} {}

\bibitem[\protect\citeauthoryear{{Claret}}{{Claret}}{2017}]{2017A&A...600A..30C}
{Claret} A.,  2017, \mn@doi [\aap] {10.1051/0004-6361/201629705}, \href {http://adsabs.harvard.edu/abs/2017A%26A...600A..30C} {600, A30}

\bibitem[\protect\citeauthoryear{{Claret} \& {Bloemen}}{{Claret} \& {Bloemen}}{2011}]{claret2011}
{Claret} A.,  {Bloemen} S.,  2011, \mn@doi [\aap] {10.1051/0004-6361/201116451}, \href {https://ui.adsabs.harvard.edu/abs/2011A&A...529A..75C} {529, A75}

\bibitem[\protect\citeauthoryear{{Collier Cameron} et~al.,}{{Collier Cameron} et~al.}{2010}]{colliercameron2010b}
{Collier Cameron} A.,  et~al., 2010, \mn@doi [\mnras] {10.1111/j.1365-2966.2010.16922.x}, \href {https://ui.adsabs.harvard.edu/abs/2010MNRAS.407..507C} {407, 507}

\bibitem[\protect\citeauthoryear{{Collins}, {Kielkopf}, {Stassun}  \& {Hessman}}{{Collins} et~al.}{2017}]{aij}
{Collins} K.~A.,  {Kielkopf} J.~F.,  {Stassun} K.~G.,   {Hessman} F.~V.,  2017, \mn@doi [\aj] {10.3847/1538-3881/153/2/77}, \href {https://ui.adsabs.harvard.edu/abs/2017AJ....153...77C} {153, 77}

\bibitem[\protect\citeauthoryear{{Cutri} et~al.,}{{Cutri} et~al.}{2003}]{2MASS}
{Cutri} R.~M.,  et~al., 2003, VizieR Online Data Catalog, \href {https://ui.adsabs.harvard.edu/abs/2003yCat.2246....0C} {p. II/246}

\bibitem[\protect\citeauthoryear{{Cutri} et~al.,}{{Cutri} et~al.}{2021a}]{allwise}
{Cutri} R.~M.,  et~al., 2021a, VizieR Online Data Catalog, \href {https://ui.adsabs.harvard.edu/abs/2014yCat.2328....0C} {p. II/328}

\bibitem[\protect\citeauthoryear{{Cutri} et~al.,}{{Cutri} et~al.}{2021b}]{WISE}
{Cutri} R.~M.,  et~al., 2021b, VizieR Online Data Catalog, \href {https://ui.adsabs.harvard.edu/abs/2014yCat.2328....0C} {p. II/328}

\bibitem[\protect\citeauthoryear{{Dawson} \& {Johnson}}{{Dawson} \& {Johnson}}{2018}]{2018DawsonJohnson}
{Dawson} R.~I.,  {Johnson} J.~A.,  2018, \mn@doi [\araa] {10.1146/annurev-astro-081817-051853}, \href {https://ui.adsabs.harvard.edu/abs/2018ARA&A..56..175D} {56, 175}

\bibitem[\protect\citeauthoryear{{Donati}, {Semel}, {Carter}, {Rees}  \& {Collier Cameron}}{{Donati} et~al.}{1997}]{donati1997}
{Donati} J.~F.,  {Semel} M.,  {Carter} B.~D.,  {Rees} D.~E.,   {Collier Cameron} A.,  1997, \mn@doi [\mnras] {10.1093/mnras/291.4.658}, \href {https://ui.adsabs.harvard.edu/abs/1997MNRAS.291..658D} {291, 658}

\bibitem[\protect\citeauthoryear{{Dotter}}{{Dotter}}{2016}]{2016ApJS..222....8D}
{Dotter} A.,  2016, \mn@doi [\apjs] {10.3847/0067-0049/222/1/8}, \href {https://ui.adsabs.harvard.edu/abs/2016ApJS..222....8D} {222, 8}

\bibitem[\protect\citeauthoryear{{Eastman}, {Gaudi}  \& {Agol}}{{Eastman} et~al.}{2013}]{eastman2013}
{Eastman} J.,  {Gaudi} B.~S.,   {Agol} E.,  2013, \mn@doi [\pasp] {10.1086/669497}, \href {https://ui.adsabs.harvard.edu/abs/2013PASP..125...83E} {125, 83}

\bibitem[\protect\citeauthoryear{F\H{u}r\'esz}{F\H{u}r\'esz}{2008}]{gaborthesis}
F\H{u}r\'esz G.,  2008, PhD thesis, University of Szeged, Hungary

\bibitem[\protect\citeauthoryear{{Fielding}, {McKee}, {Socrates}, {Cunningham}  \& {Klein}}{{Fielding} et~al.}{2015}]{fielding2015}
{Fielding} D.~B.,  {McKee} C.~F.,  {Socrates} A.,  {Cunningham} A.~J.,   {Klein} R.~I.,  2015, \mn@doi [\mnras] {10.1093/mnras/stv836}, \href {https://ui.adsabs.harvard.edu/abs/2015MNRAS.450.3306F} {450, 3306}

\bibitem[\protect\citeauthoryear{{Foucart} \& {Lai}}{{Foucart} \& {Lai}}{2011}]{foucart2011}
{Foucart} F.,  {Lai} D.,  2011, \mn@doi [\mnras] {10.1111/j.1365-2966.2010.18176.x}, \href {https://ui.adsabs.harvard.edu/abs/2011MNRAS.412.2799F} {412, 2799}

\bibitem[\protect\citeauthoryear{{Gaia Collaboration} et~al.,}{{Gaia Collaboration} et~al.}{2022a}]{Gaiadr3}
{Gaia Collaboration} et~al., 2022a, \mn@doi [arXiv e-prints] {10.48550/arXiv.2208.00211}, \href {https://ui.adsabs.harvard.edu/abs/2022arXiv220800211G} {p. arXiv:2208.00211}

\bibitem[\protect\citeauthoryear{{Gaia Collaboration} et~al.,}{{Gaia Collaboration} et~al.}{2022b}]{2022arXiv220800211G}
{Gaia Collaboration} et~al., 2022b, \mn@doi [arXiv e-prints] {10.48550/arXiv.2208.00211}, \href {https://ui.adsabs.harvard.edu/abs/2022arXiv220800211G} {p. arXiv:2208.00211}

\bibitem[\protect\citeauthoryear{{Garcia-Mejia}, {Charbonneau}, {Fabricant}, {Irwin}, {Fata}, {Zajac}  \& {Doherty}}{{Garcia-Mejia} et~al.}{2020}]{tierras}
{Garcia-Mejia} J.,  {Charbonneau} D.,  {Fabricant} D.,  {Irwin} J.,  {Fata} R.,  {Zajac} J.~M.,   {Doherty} P.~E.,  2020, in {Marshall} H.~K.,  {Spyromilio} J.,   {Usuda} T.,  eds,  Society of Photo-Optical Instrumentation Engineers (SPIE) Conference Series Vol. 11445, Ground-based and Airborne Telescopes VIII. p. 114457R (\mn@eprint {arXiv} {2012.09744}), \mn@doi{10.1117/12.2561467}

\bibitem[\protect\citeauthoryear{{Gray} \& {Corbally}}{{Gray} \& {Corbally}}{1994}]{1994AJ....107..742G}
{Gray} R.~O.,  {Corbally} C.~J.,  1994, \mn@doi [\aj] {10.1086/116893}, \href {http://adsabs.harvard.edu/abs/1994AJ....107..742G} {107, 742}

\bibitem[\protect\citeauthoryear{{Guerrero} et~al.,}{{Guerrero} et~al.}{2021}]{Guerrero2021}
{Guerrero} N.~M.,  et~al., 2021, \mn@doi [\apjs] {10.3847/1538-4365/abefe1}, \href {https://ui.adsabs.harvard.edu/abs/2021ApJS..254...39G} {254, 39}

\bibitem[\protect\citeauthoryear{{Heller}}{{Heller}}{1993}]{1993ApJ...408..337H}
{Heller} C.~H.,  1993, \mn@doi [\apj] {10.1086/172591}, \href {https://ui.adsabs.harvard.edu/abs/1993ApJ...408..337H} {408, 337}

\bibitem[\protect\citeauthoryear{{H{\o}g} et~al.,}{{H{\o}g} et~al.}{2000}]{tycho}
{H{\o}g} E.,  et~al., 2000, \aap, \href {https://ui.adsabs.harvard.edu/abs/2000A&A...355L..27H} {355, L27}

\bibitem[\protect\citeauthoryear{{Howell}, {Everett}, {Sherry}, {Horch}  \& {Ciardi}}{{Howell} et~al.}{2011}]{speckle2011}
{Howell} S.~B.,  {Everett} M.~E.,  {Sherry} W.,  {Horch} E.,   {Ciardi} D.~R.,  2011, \mn@doi [\aj] {10.1088/0004-6256/142/1/19}, \href {https://ui.adsabs.harvard.edu/abs/2011AJ....142...19H} {142, 19}

\bibitem[\protect\citeauthoryear{{Huang} et~al.,}{{Huang} et~al.}{2020}]{Huang2020}
{Huang} C.~X.,  et~al., 2020, \mn@doi [Research Notes of the American Astronomical Society] {10.3847/2515-5172/abca2e}, \href {https://ui.adsabs.harvard.edu/abs/2020RNAAS...4..204H} {4, 204}

\bibitem[\protect\citeauthoryear{{Irwin}, {Berta-Thompson}, {Charbonneau}, {Dittmann}, {Falco}, {Newton}  \& {Nutzman}}{{Irwin} et~al.}{2015}]{irwin2015}
{Irwin} J.~M.,  {Berta-Thompson} Z.~K.,  {Charbonneau} D.,  {Dittmann} J.,  {Falco} E.~E.,  {Newton} E.~R.,   {Nutzman} P.,  2015, in 18th Cambridge Workshop on Cool Stars, Stellar Systems, and the Sun. pp 767--772 (\mn@eprint {arXiv} {1409.0891}), \mn@doi{10.48550/arXiv.1409.0891}

\bibitem[\protect\citeauthoryear{{Jenkins}}{{Jenkins}}{2002}]{Jenkins2002}
{Jenkins} J.~M.,  2002, \mn@doi [\apj] {10.1086/341136}, \href {https://ui.adsabs.harvard.edu/abs/2002ApJ...575..493J} {575, 493}

\bibitem[\protect\citeauthoryear{{Jenkins} et~al.,}{{Jenkins} et~al.}{2010}]{Jenkins2010}
{Jenkins} J.~M.,  et~al., 2010, in {Radziwill} N.~M.,  {Bridger} A.,  eds,  Society of Photo-Optical Instrumentation Engineers (SPIE) Conference Series Vol. 7740, Software and Cyberinfrastructure for Astronomy. p. 77400D, \mn@doi{10.1117/12.856764}

\bibitem[\protect\citeauthoryear{{Jenkins} et~al.,}{{Jenkins} et~al.}{2016}]{jenkins2016}
{Jenkins} J.~M.,  et~al., 2016, in {Chiozzi} G.,  {Guzman} J.~C.,  eds,  Society of Photo-Optical Instrumentation Engineers (SPIE) Conference Series Vol. 9913, Software and Cyberinfrastructure for Astronomy IV. p. 99133E, \mn@doi{10.1117/12.2233418}

\bibitem[\protect\citeauthoryear{{Jenkins}, {Tenenbaum}, {Seader}, {Burke}, {McCauliff}, {Smith}, {Twicken}  \& {Chandrasekaran}}{{Jenkins} et~al.}{2020}]{Jenkins2020}
{Jenkins} J.~M.,  {Tenenbaum} P.,  {Seader} S.,  {Burke} C.~J.,  {McCauliff} S.~D.,  {Smith} J.~C.,  {Twicken} J.~D.,   {Chandrasekaran} H.,  2020, {Kepler Data Processing Handbook: Transiting Planet Search}, Kepler Science Document KSCI-19081-003, id. 9. Edited by Jon M. Jenkins.

\bibitem[\protect\citeauthoryear{{Jensen}}{{Jensen}}{2013}]{jensen2013}
{Jensen} E.,  2013, {Tapir: A web interface for transit/eclipse observability}, Astrophysics Source Code Library, record ascl:1306.007 (\mn@eprint {ascl} {1306.007})

\bibitem[\protect\citeauthoryear{{Johnson}, {Cochran}, {Addison}, {Tinney}  \& {Wright}}{{Johnson} et~al.}{2017}]{2017AJ....154..137J}
{Johnson} M.~C.,  {Cochran} W.~D.,  {Addison} B.~C.,  {Tinney} C.~G.,   {Wright} D.~J.,  2017, \mn@doi [\aj] {10.3847/1538-3881/aa8462}, \href {https://ui.adsabs.harvard.edu/abs/2017AJ....154..137J} {154, 137}

\bibitem[\protect\citeauthoryear{{Koposov}}{{Koposov}}{2020}]{2020zndo...4002972K}
{Koposov} S.,  2020, {segasai/minimint: First official release}, Zenodo, \mn@doi{10.5281/zenodo.4002972}

\bibitem[\protect\citeauthoryear{{Kov{\'a}cs}, {Zucker}  \& {Mazeh}}{{Kov{\'a}cs} et~al.}{2002}]{Kovacs2002}
{Kov{\'a}cs} G.,  {Zucker} S.,   {Mazeh} T.,  2002, \mn@doi [\aap] {10.1051/0004-6361:20020802}, \href {https://ui.adsabs.harvard.edu/abs/2002A&A...391..369K} {391, 369}

\bibitem[\protect\citeauthoryear{{Kraft}}{{Kraft}}{1967}]{kraftbreak1967}
{Kraft} R.~P.,  1967, \mn@doi [\apj] {10.1086/149359}, \href {https://ui.adsabs.harvard.edu/abs/1967ApJ...150..551K} {150, 551}

\bibitem[\protect\citeauthoryear{{Kreidberg}}{{Kreidberg}}{2015}]{2015PASP..127.1161K}
{Kreidberg} L.,  2015, \mn@doi [\pasp] {10.1086/683602}, \href {http://adsabs.harvard.edu/abs/2015PASP..127.1161K} {127, 1161}

\bibitem[\protect\citeauthoryear{{Kuffmeier}, {Dullemond}, {Reissl}  \& {Goicovic}}{{Kuffmeier} et~al.}{2021}]{kuffmeier2021}
{Kuffmeier} M.,  {Dullemond} C.~P.,  {Reissl} S.,   {Goicovic} F.~G.,  2021, \mn@doi [\aap] {10.1051/0004-6361/202039614}, \href {https://ui.adsabs.harvard.edu/abs/2021A&A...656A.161K} {656, A161}

\bibitem[\protect\citeauthoryear{{Kunimoto}, {Tey}, {Fong}, {Hesse}, {Shporer}, {Fausnaugh}, {Vanderspek}  \& {Ricker}}{{Kunimoto} et~al.}{2022}]{Kunimoto2022}
{Kunimoto} M.,  {Tey} E.,  {Fong} W.,  {Hesse} K.,  {Shporer} A.,  {Fausnaugh} M.,  {Vanderspek} R.,   {Ricker} G.,  2022, \mn@doi [Research Notes of the American Astronomical Society] {10.3847/2515-5172/aca158}, \href {https://ui.adsabs.harvard.edu/abs/2022RNAAS...6..236K} {6, 236}

\bibitem[\protect\citeauthoryear{{Kurucz}}{{Kurucz}}{1992}]{kurucz1992}
{Kurucz} R.~L.,  1992, in {Barbuy} B.,  {Renzini} A.,  eds, ~ Vol. 149, The Stellar Populations of Galaxies. p.~225

\bibitem[\protect\citeauthoryear{{Lai}}{{Lai}}{2012}]{lai2012}
{Lai} D.,  2012, \mn@doi [\mnras] {10.1111/j.1365-2966.2012.20893.x}, \href {https://ui.adsabs.harvard.edu/abs/2012MNRAS.423..486L} {423, 486}

\bibitem[\protect\citeauthoryear{{Li}, {Tenenbaum}, {Twicken}, {Burke}, {Jenkins}, {Quintana}, {Rowe}  \& {Seader}}{{Li} et~al.}{2019}]{Li2019}
{Li} J.,  {Tenenbaum} P.,  {Twicken} J.~D.,  {Burke} C.~J.,  {Jenkins} J.~M.,  {Quintana} E.~V.,  {Rowe} J.~F.,   {Seader} S.~E.,  2019, \mn@doi [\pasp] {10.1088/1538-3873/aaf44d}, \href {https://ui.adsabs.harvard.edu/abs/2019PASP..131b4506L} {131, 024506}

\bibitem[\protect\citeauthoryear{{Lightkurve Collaboration} et~al.,}{{Lightkurve Collaboration} et~al.}{2018}]{lightkurve}
{Lightkurve Collaboration} et~al., 2018, {Lightkurve: Kepler and TESS time series analysis in Python}, Astrophysics Source Code Library (\mn@eprint {ascl} {1812.013})

\bibitem[\protect\citeauthoryear{{Lin}, {Bodenheimer}  \& {Richardson}}{{Lin} et~al.}{1996}]{1996Natur.380..606L}
{Lin} D.~N.~C.,  {Bodenheimer} P.,   {Richardson} D.~C.,  1996, \mn@doi [\nat] {10.1038/380606a0}, \href {https://ui.adsabs.harvard.edu/abs/1996Natur.380..606L} {380, 606}

\bibitem[\protect\citeauthoryear{{Lubow} \& {Ogilvie}}{{Lubow} \& {Ogilvie}}{2000}]{lubow2000}
{Lubow} S.~H.,  {Ogilvie} G.~I.,  2000, \mn@doi [\apj] {10.1086/309101}, \href {https://ui.adsabs.harvard.edu/abs/2000ApJ...538..326L} {538, 326}

\bibitem[\protect\citeauthoryear{{Mandel} \& {Agol}}{{Mandel} \& {Agol}}{2002}]{mandelAgol2002}
{Mandel} K.,  {Agol} E.,  2002, \mn@doi [\apjl] {10.1086/345520}, \href {https://ui.adsabs.harvard.edu/abs/2002ApJ...580L.171M} {580, L171}

\bibitem[\protect\citeauthoryear{{Masuda} \& {Winn}}{{Masuda} \& {Winn}}{2020}]{2020AJ....159...81M}
{Masuda} K.,  {Winn} J.~N.,  2020, \mn@doi [\aj] {10.3847/1538-3881/ab65be}, \href {https://ui.adsabs.harvard.edu/abs/2020AJ....159...81M} {159, 81}

\bibitem[\protect\citeauthoryear{{Matsakos} \& {K{\"o}nigl}}{{Matsakos} \& {K{\"o}nigl}}{2017}]{matsakos2017}
{Matsakos} T.,  {K{\"o}nigl} A.,  2017, \mn@doi [\aj] {10.3847/1538-3881/153/2/60}, \href {https://ui.adsabs.harvard.edu/abs/2017AJ....153...60M} {153, 60}

\bibitem[\protect\citeauthoryear{{McCully}, {Daily}, {Brandt}, {Johnson}, {Bowman}  \& {Harbeck}}{{McCully} et~al.}{2022}]{McCully22}
{McCully} C.,  {Daily} M.,  {Brandt} G.~M.,  {Johnson} M.~C.,  {Bowman} M.,   {Harbeck} D.-R.,  2022, in Software and Cyberinfrastructure for Astronomy VII. p. 1218914 (\mn@eprint {arXiv} {2212.11381}), \mn@doi{10.1117/12.2630667}

\bibitem[\protect\citeauthoryear{{McLaughlin}}{{McLaughlin}}{1924}]{mclaughlin}
{McLaughlin} D.~B.,  1924, \mn@doi [\apj] {10.1086/142826}, \href {https://ui.adsabs.harvard.edu/abs/1924ApJ....60...22M} {60, 22}

\bibitem[\protect\citeauthoryear{{Perryman} et~al.,}{{Perryman} et~al.}{1997}]{1997AA...323L..49P}
{Perryman} M.~A.~C.,  et~al., 1997, \aap, \href {https://ui.adsabs.harvard.edu/abs/1997A&A...323L..49P} {500, 501}

\bibitem[\protect\citeauthoryear{{Pont} et~al.,}{{Pont} et~al.}{2009}]{HD806062009Pont}
{Pont} F.,  et~al., 2009, \mn@doi [\aap] {10.1051/0004-6361/200912463}, \href {https://ui.adsabs.harvard.edu/abs/2009A&A...502..695P} {502, 695}

\bibitem[\protect\citeauthoryear{{Queloz} et~al.,}{{Queloz} et~al.}{2010}]{WASP8_2010Queloz}
{Queloz} D.,  et~al., 2010, \mn@doi [\aap] {10.1051/0004-6361/201014768}, \href {https://ui.adsabs.harvard.edu/abs/2010A&A...517L...1Q} {517, L1}

\bibitem[\protect\citeauthoryear{{Rasio} \& {Ford}}{{Rasio} \& {Ford}}{1996}]{1996Sci...274..954R}
{Rasio} F.~A.,  {Ford} E.~B.,  1996, \mn@doi [Science] {10.1126/science.274.5289.954}, \href {https://ui.adsabs.harvard.edu/abs/1996Sci...274..954R} {274, 954}

\bibitem[\protect\citeauthoryear{{Ricker} et~al.,}{{Ricker} et~al.}{2015}]{ricker2015}
{Ricker} G.~R.,  et~al., 2015, \mn@doi [Journal of Astronomical Telescopes, Instruments, and Systems] {10.1117/1.JATIS.1.1.014003}, \href {https://ui.adsabs.harvard.edu/abs/2015JATIS...1a4003R} {1, 014003}

\bibitem[\protect\citeauthoryear{{Rogers}, {Lin}  \& {Lau}}{{Rogers} et~al.}{2012}]{2012ApJ...758L...6R}
{Rogers} T.~M.,  {Lin} D.~N.~C.,   {Lau} H.~H.~B.,  2012, \mn@doi [\apjl] {10.1088/2041-8205/758/1/L6}, \href {https://ui.adsabs.harvard.edu/abs/2012ApJ...758L...6R} {758, L6}

\bibitem[\protect\citeauthoryear{{Rossiter}}{{Rossiter}}{1924}]{rossiter}
{Rossiter} R.~A.,  1924, \mn@doi [\apj] {10.1086/142825}, \href {https://ui.adsabs.harvard.edu/abs/1924ApJ....60...15R} {60, 15}

\bibitem[\protect\citeauthoryear{{Santerne} et~al.,}{{Santerne} et~al.}{2014}]{Kepler420_2014Santerne}
{Santerne} A.,  et~al., 2014, \mn@doi [\aap] {10.1051/0004-6361/201424158}, \href {https://ui.adsabs.harvard.edu/abs/2014A&A...571A..37S} {571, A37}

\bibitem[\protect\citeauthoryear{{Scott}, {Howell}, {Horch}  \& {Everett}}{{Scott} et~al.}{2018}]{NESSI}
{Scott} N.~J.,  {Howell} S.~B.,  {Horch} E.~P.,   {Everett} M.~E.,  2018, \mn@doi [\pasp] {10.1088/1538-3873/aab484}, \href {https://ui.adsabs.harvard.edu/abs/2018PASP..130e4502S} {130, 054502}

\bibitem[\protect\citeauthoryear{{Sikora}, {Wade}  \& {Rowe}}{{Sikora} et~al.}{2020}]{2020MNRAS.498.2456S}
{Sikora} J.,  {Wade} G.~A.,   {Rowe} J.,  2020, \mn@doi [\mnras] {10.1093/mnras/staa2444}, \href {https://ui.adsabs.harvard.edu/abs/2020MNRAS.498.2456S} {498, 2456}

\bibitem[\protect\citeauthoryear{{Siverd} et~al.,}{{Siverd} et~al.}{2018}]{Siverd18}
{Siverd} R.~J.,  et~al., 2018, in {Evans} C.~J.,  {Simard} L.,   {Takami} H.,  eds,  Society of Photo-Optical Instrumentation Engineers (SPIE) Conference Series Vol. 10702, Ground-based and Airborne Instrumentation for Astronomy VII. p. 107026C, \mn@doi{10.1117/12.2312800}

\bibitem[\protect\citeauthoryear{{Skrutskie} et~al.,}{{Skrutskie} et~al.}{2006}]{2006AJ....131.1163S}
{Skrutskie} M.~F.,  et~al., 2006, \mn@doi [\aj] {10.1086/498708}, \href {http://adsabs.harvard.edu/abs/2006AJ....131.1163S} {131, 1163}

\bibitem[\protect\citeauthoryear{{Smith} et~al.,}{{Smith} et~al.}{2012}]{2012PASP..124.1000S}
{Smith} J.~C.,  et~al., 2012, \mn@doi [\pasp] {10.1086/667697}, \href {http://adsabs.harvard.edu/abs/2012PASP..124.1000S} {124, 1000}

\bibitem[\protect\citeauthoryear{{Stassun} et~al.,}{{Stassun} et~al.}{2019}]{stassun2019}
{Stassun} K.~G.,  et~al., 2019, \mn@doi [\aj] {10.3847/1538-3881/ab3467}, \href {https://ui.adsabs.harvard.edu/abs/2019AJ....158..138S} {158, 138}

\bibitem[\protect\citeauthoryear{{Stumpe} et~al.,}{{Stumpe} et~al.}{2012}]{2012PASP..124..985S}
{Stumpe} M.~C.,  et~al., 2012, \mn@doi [\pasp] {10.1086/667698}, \href {https://ui.adsabs.harvard.edu/abs/2012PASP..124..985S} {124, 985}

\bibitem[\protect\citeauthoryear{{Stumpe}, {Smith}, {Catanzarite}, {Van Cleve}, {Jenkins}, {Twicken}  \& {Girouard}}{{Stumpe} et~al.}{2014}]{2014PASP..126..100S}
{Stumpe} M.~C.,  {Smith} J.~C.,  {Catanzarite} J.~H.,  {Van Cleve} J.~E.,  {Jenkins} J.~M.,  {Twicken} J.~D.,   {Girouard} F.~R.,  2014, \mn@doi [\pasp] {10.1086/674989}, \href {http://adsabs.harvard.edu/abs/2014PASP..126..100S} {126, 100}

\bibitem[\protect\citeauthoryear{{Thies}, {Kroupa}, {Goodwin}, {Stamatellos}  \& {Whitworth}}{{Thies} et~al.}{2011}]{thies2011}
{Thies} I.,  {Kroupa} P.,  {Goodwin} S.~P.,  {Stamatellos} D.,   {Whitworth} A.~P.,  2011, \mn@doi [\mnras] {10.1111/j.1365-2966.2011.19390.x}, \href {https://ui.adsabs.harvard.edu/abs/2011MNRAS.417.1817T} {417, 1817}

\bibitem[\protect\citeauthoryear{{Twicken} et~al.,}{{Twicken} et~al.}{2018}]{Twicken2018}
{Twicken} J.~D.,  et~al., 2018, \mn@doi [\pasp] {10.1088/1538-3873/aab694}, \href {https://ui.adsabs.harvard.edu/abs/2018PASP..130f4502T} {130, 064502}

\bibitem[\protect\citeauthoryear{{Wijnen}, {Pelupessy}, {Pols}  \& {Portegies Zwart}}{{Wijnen} et~al.}{2017}]{2017A&A...604A..88W}
{Wijnen} T.~P.~G.,  {Pelupessy} F.~I.,  {Pols} O.~R.,   {Portegies Zwart} S.,  2017, \mn@doi [\aap] {10.1051/0004-6361/201730793}, \href {https://ui.adsabs.harvard.edu/abs/2017A&A...604A..88W} {604, A88}

\bibitem[\protect\citeauthoryear{{Winn} et~al.,}{{Winn} et~al.}{2009}]{HD806062009Winn}
{Winn} J.~N.,  et~al., 2009, \mn@doi [\apj] {10.1088/0004-637X/703/2/2091}, \href {https://ui.adsabs.harvard.edu/abs/2009ApJ...703.2091W} {703, 2091}

\bibitem[\protect\citeauthoryear{{Zhou}, {Latham}, {Bieryla}, {Beatty}, {Buchhave}, {Esquerdo}, {Berlind}  \& {Calkins}}{{Zhou} et~al.}{2016}]{zhou2016}
{Zhou} G.,  {Latham} D.~W.,  {Bieryla} A.,  {Beatty} T.~G.,  {Buchhave} L.~A.,  {Esquerdo} G.~A.,  {Berlind} P.,   {Calkins} M.~L.,  2016, \mn@doi [\mnras] {10.1093/mnras/stw1107}, \href {https://ui.adsabs.harvard.edu/abs/2016MNRAS.460.3376Z} {460, 3376}

\bibitem[\protect\citeauthoryear{{Zhou} et~al.,}{{Zhou} et~al.}{2019}]{2019AJ....157...31Z}
{Zhou} G.,  et~al., 2019, \mn@doi [\aj] {10.3847/1538-3881/aaf1bb}, \href {https://ui.adsabs.harvard.edu/abs/2019AJ....157...31Z} {157, 31}

\makeatother
\end{thebibliography}
